\documentclass[referee]{raa}
\usepackage{txfonts}
\usepackage{graphicx,subfigure}
\usepackage[colorlinks,linkcolor=blue,urlcolor=blue,citecolor=blue]{hyperref}
 \usepackage{natbib}
 \usepackage{mathrsfs}
 \usepackage{ulem}
 \usepackage{verbatim}
 \usepackage{lineno}
 \usepackage{ulem}
 \usepackage{gensymb}
\usepackage[margin=2cm]{geometry}
\newcommand{\kms}{km\ s$^{-1}$}

\begin{document}
% \linenumbers
\title{The Relative Calibration of Radial Velocity for LAMOST Medium Resolution Stellar Spectra}
\author{Jianping Xiong\inst{1,2},
        Bo Zhang\inst{3}, 
        Chao Liu\thanks{corresponding author: liuchao@nao.cas.cn}\inst{4,2}, 
        Jiao Li\inst{4,5}, 
        Yongheng Zhao\inst{1,2,7},
        Yonghui Hou\inst{6,7}
   }
\institute{
        Key Laboratory of Optical Astronomy, National Astronomical Observatories, Chinese Academy of Sciences, Beijing, 100101, China\\ 
        \and
        University of Chinese Academy of Sciences,Beijing, 100049, China\\
        \and
        Department of Astronomy, Beijing Normal University, Beijing, 100875, China\\
        \and 
        Key Laboratory of Space Astronomy and Technology, National Astronomical Observatories, Chinese Academy of Sciences,  Beijing, 100101, China\\ 
        \and 
        Yunnan observatories, Chinese Academy of Sciences, Kunming, 650011, China\\
        \and
        Nanjing Institute of Astronomical Optics, \& Technology, National Astronomical Observatories, Chinese Academy of Sciences, Nanjing 210042, China\\
        \and
        School of Astronomy and Space Science, University of Chinese Academy of Sciences, Beijing, 100049, China
}

\abstract{The Large Sky Area Multi-Object Fiber Spectroscopic Telescope (LAMOST) started median-resolution spectroscopic (MRS, R$\sim$7500) survey since October 2018. The main scientific goals of MRS, including binary stars, pulsators, and other variable stars are launched with a time-domain spectroscopic survey. However, the systematic errors, including the bias induced from wavelength calibration and the systematic difference between different spectrographs have to be carefully considered during radial velocity measurement. In this work, we provide a technique to correct the systematics in the wavelength calibration based on the relative radial velocity measurements from LAMOST MRS spectra. We show that, for the stars with multi-epoch spectra, the systematic bias which is induced from the exposures of different nights can be well corrected for LAMOST MRS in each spectrograph. And the precision of radial velocity zero-point of multi-epoch time-domain observations reaches below 0.5\,\kms. ~As a by-product, we also give the constant star candidates\footnote{The constant stars here in this work represent for the stars with low radial velocity variations. They are not necessarily absolute constant stars, but may not show significant 
radial velocity variation during the time-domain survey}, which can be the secondary radial-velocity standard star candidates of LAMOST MRS time-domain surveys. 
}

\authorrunning{J.-P. Xiong et al. }      
\titlerunning{RV Relative Calibration For LAMOST-MRS }
\maketitle

\section{Introduction}

Over the past years, the large spectroscopic surveys, such as  RAVE~\citep{2006AJ....132.1645S, 2020AJ....160...83S}, SDSS/SEGUE~\citep{2009AJ....137.4377Y}, LAMOST~\citep{2012RAA....12.1197C, 2012RAA....12..735D, 2012RAA....12..723Z, 2015RAA....15.1095L}, APOGEE~\citep{2017AJ....154...94M}, GALAH~\citep{2015MNRAS.449.2604D}, Gaia~\citep{2012Msngr.147...25G, 2004MNRAS.354.1223K, 2018A&A...616A...5C} have yielded a large number of stellar spectra to understand the galaxy formation and evolution~\citep{2015RAA....15.2204G, 2021MNRAS.501.1046M}.
As a fundamental property that can be measured directly from stellar spectra, radial velocity (RV) is the basic ingredient for many studies, such as the study on the properties of binary~\citep{2014ApJ...788L..37G, 2018RAA....18...52T, 2012Sci...337..444S, 2013A&A...550A.107S}, Milky Way dynamics~\citep{2020ApJ...899..110T}, asteroseismology~\citep{2019A&A...622A.190A}, and even searching for the black holes~\citep{2019Natur.575..618L}.

For large spectroscopic surveys, the radial velocity uncertainty of low-resolution spectra is larger $3-5$\,\kms for SEGUE in R$\sim$1800~\citep{2009AJ....137.4377Y}, and around $3-5$\,\kms for LAMOST in R$\sim$1800,~\citep{2014IAUS..306..340W} .   Such precision is very useful in the Milky Way studies, however, it is not enough in lots of studies of stars and stellar systems, e.g., the internal dynamics of open clusters (with the radial velocity dispersion of $<1$\,\kms ~\citep{2021ApJ...912..162P}), the disruption process of open clusters (with the radial velocity dispersion around $1-3$\,\kms ~\citep{2021arXiv210607658P}), the binary stars with a larger range of period, the low-amplitude variables. The precision of radial velocity is required down to $<1$\,\kms for these studies. This is essentially the motivation of the LAMOST medium-resolution spectroscopic (MRS) with $R\sim7500$.

Since October 2018, LAMOST has started the 5-year MRS survey~\citep{2020arXiv200507210L}. A large amount of and medium resolution stellar spectra will be obtained. It provides an opportunity to study the kinematic and dynamic of stars more accurately. For LAMOST MRS survey,~\cite{2019arXiv190804773W} derived radial velocity with cross-correlation method, and the precision achieves to 1.36~\,\kms,\ 1.08~\,\kms,\ 0.91~\,\kms\ for the spectra at the single-to-noise ratio (SNR) of 10, 20, 50,  respectively, after calibrated with the radial velocity standard stars~\citep{2018AJ....156...90H}.

However, there are a few issues need to  be well addressed. One is that the wavelength calibration of the spectra of LAMOST demonstrates that it may shift by a few \kms\ during one night, which is likely due to the variation of the environments of the spectrographs (Zhang Hao-Tong priv. comm.). The systematic shift between different exposures may severely affect the orbital parameter estimates in binary studies. In the mean time, because LAMOST contains 16 spectrographs, one star can be observed by different spectrographs in different nights, depending on the specific fiber assignment strategy. This may bring another systematic difference in time-domain spectra, due to the systematically different wavelength calibration between different spectrographs. It means that the instability of the wavelength calibration may induce a large fraction of systematic errors in the uncertainty of radial velocity. 

It is noted that many studies, e.g. the pulsation of stars, the binarities, etc., do not require absolute radial velocities, but only relative one. Specifically, at time $t$, a star has a directly measured radial velocity such as $v_t=v_{0}+\Delta v_{t}+\epsilon$, where $v_0$ is the systematic radial velocity of the star, $\Delta v_{t}$ is the term of possible variation if the star is a pulsator or a companion in a binary system. And $\epsilon$ is the measurement uncertainty. To understand the pulsational process or the parameters of the orbits of a binary system using a series time-domain radial velocities, $\Delta v_{t}$ is sufficient rather than using $v_{t}$.

Therefore, the goal of this work is two-fold: first we correct the systematic bias in radial velocity; second, we provide relative radial velocity measurement for the 92,342 time-domain MRS stars. As a by-product, we give the constant star candidates (The constant stars here in this work represent for the stars with low radial velocity variations. They are not necessarily absolute constant stars, but may not show significant 
radial velocity variation during the time-domain survey), which do not show significant radial velocity shift in the MRS time-domain spectra.

This paper is organized as follows. The LAMOST MRS and the data are described in Section~\ref{sec:data}. The method of relative calibration is described in Section~\ref{sec:method}. The results are indicated in Section~\ref{sec:result}. The special cases are analyzed in Section~\ref{sec:Discussion}. Finally, we summarize in Section~\ref{sec:Conclusion}.

\section{Data} \label{sec:data}

Guoshoujing Telescope (also known as LAMOST) is a 4-meter class reflective Schmidt telescope with 5-degree field-of-view. Totally 4000 fibers are installed at the 1.75\,m-diameter large focal plane. 16 spectrographs, each of which accepts 250 fibers, are used to take spectra simultaneously. Each spectrograph contains two arms. For medium spectroscopic observation, the blue arm covers 4950-5350\,\AA\ and the red arm covers 6300-6800\,\AA\ with $R\sim 7500$.

As mentioned in~\cite{2020arXiv200507210L}, the LAMOST MRS take observations in 14 bright and gray nights per month. About 2000 square degree sky area, including various Galactic latitudes, are designed for the time-domain spectroscopic survey. It is expected that, after a 5-year MRS survey, there will be around 200\,000 stars with $G<14$\,mag\footnote{$G$ is the band without filter in \emph{Gaia} survey~\citep{2016A&A...595A...1G}.} to be observed multiple times.

The seeing, cloud coverage and checking of polluted light are evaluated to remove the low-quality observations from two-dimensional(2D) frames at first. Then the 2D medium-resolution frames are processed by LAMOST 2D pipeline, which includes bias subtraction, fiber tracing, wavelength calibration, spectral extraction and sky subtraction. For an observation target, the spectra from the same night will be merged into a coadded spectrum, and the separate spectrum of each exposure is provided as well.

\section{Method} 
\label{sec:method}

The directly measured radial velocity ($\widetilde{v}_{i,t}$) of a star $i$ observed at time $t$ is composed of the intrinsic radial velocity ($v_{i,t}$), the systematic error ($\Delta u_{t}$) due to the wavelength calibration during the observation and the measurement uncertainty ($\epsilon_{i,t}$) induced during measurement, i.e.
\begin{equation}\label{eq:rv}
     \widetilde{v}_{i,t} = v_{i,t}-\Delta u_{t}+\epsilon_{i,t}.
\end{equation}
In general, $v_{i,t}$ is independent on $t$. For a constant star, its radial velocity does not change with observation time (i.e. $v_{i,t_{1}}=v_{i,t_{2}}$). When star $i$ is in a binary system or is a variable, $v_{i,t}$ changes periodically with $t$. For the former case, the relative radial velocities between two exposures should be around zero, while for the latter case, the differential radial velocities are sufficient for solving out the velocity curve with higher precision than that derived from the absolutely calibrated radial velocities. Therefore, in this work, we concentrate on the relative radial velocity only.

\subsection{Relative radial velocity with maximum likelihood}

We develop a likelihood method to measure the relative radial velocity for a star from its medium-resolution spectra. First, each spectrum observed by LAMOST has to be normalized. We normalize each spectrum by dividing by its pseudo-continuum, the pseudo-continuum is calculated by the follows: each spectrum is fitted with a smoothing spline function iteratively, then the pixels with a value greater than the median value by 3 times of standard deviation are eliminated in each segmented wavelength (e.g. 100\AA). And the iterations are 3 \citep{2021arXiv210511624Z}.

Then, for the $i$th star with multiple exposures, we select the spectrum, denoted as $f_{ref,i}(\lambda)$, with highest signal-to-noise ratio as the reference for the star. For the spectrum $f_{t,i}(\lambda, v)$, which is the spectrum of $i$th star observed at time $t$, the likelihood distribution of the relative radial velocity $v$ is 
\begin{equation}\label{eq:likelihood}
p(v)=\prod_{\lambda}\frac{\exp[-\frac{(f_{t,i}(\lambda,v)-f_{ref,i}(\lambda))^2}{2(\sigma_{t,i}^2+\sigma_{ref,i}^2)}]}
{\sqrt{2\pi(\sigma_{t,i}^2+\sigma_{ref,i}^2)}},
\end{equation}
In which $\sigma_{t,i}$ and $\sigma_{ref,i}$ are the observational errors from the observed and reference spectra, respectively. In order to avoid the arithmetic errors caused by minimal values, the logarithmic form as written in below is actually used:
\begin{equation}\label{eq:lnP}
    \ln p(v)=\sum_{\lambda}[-\frac{(f_{t,i}(\lambda,v)-f_{ref,i}(\lambda))^2}
      {2(\sigma_{t,i}^2+\sigma_{ref,i}^2)}-0.5\ln(\sigma_{t,i}^2+\sigma_{ref,i}^2)-0.5\ln2\pi]
\end{equation}

\subsection{Calibrate systematic bias in radial velocity}

In general, the wavelength calibration is based on the arc lamp, such as Fe-Ar lamps etc. However, according to the tests, the wavelength calibration may be suffered from a few \kms\ systematic shift when comparing between different observations. In addition, the wavelength calibration of the red arm is more precise than the blue arm for LAMOST. In a few observation fields, radial velocities of constant stars obtained from high-resolution spectra can be used to further calibrate the wavelength in absolute sense, since the precision of these radial velocities can be as high as a few hundreds m s$^{-1}$ (see~\citealt{2018AJ....156...90H}). However, there is no such high-resolution constant stars been identified in every LAMOST observation field. Therefore, we can not only rely on the identified constant stars to improve the velocity accuracy but require an approach in the manner of self-calibration.

Before we start the calibration process, we assume that the relative radial velocities of constant stars should be zero. The calibration process contains two steps in each iteration. 

As the first step, we consider each star $i$. We calculate its relative radial velocity variation over all exposures such as
\begin{equation}
    \sigma_i^2 = \frac{1}{m}\sum_{t=1}^{m}{(v_{i,t}-\bar{v_i})^2},
\end{equation}
where $v_{i,t}$ is the relative radial velocity of star $i$ in the $t$th observation, $\bar{v_i}$ the mean relative velocity over all exposures, and $m$ the number of exposures. For all stars observed by a spectrograph, the mean radial velocity variation is $\bar{\sigma}=\sum_{i}{\sigma_i}/n$. The stars with $\sigma_i<\bar{\sigma}$ are selected as the candidate constant stars.

In the second step, we consider the velocities of stars in each exposure. We derive the mean relative velocities, denoted as $\bar{v_t}$, of the candidate constant stars selected from the first step in the $t$th exposure. Because the ground true relative radial velocity of constant stars should be zero, the non-zero $\bar{v_t}$ is mainly contributed by the systematic bias in wavelength calibration. Therefore,  $\bar{v_t}$ is subtracted from $v_{i,t}$ for each star, including constant and non-constant stars. After subtraction, the candidate constant stars should be more concentrated around zero.

Note that some contaminators may be induced in the first step, we go back to step one and select the new candidate constant stars again with corrected relative radial velocities.

After a few iterations with the above two steps, when $\bar{v_t}$ tends to be very close to zero, and the variation of $\bar{v_t}$ for each exposure is lower than a predefined small threshold (we adopt any of $\vert \bar{v_t}-\bar{v}_{t-1} \vert <0.01$ \kms\ ), the iteration is stopped.

\subsection{Validation with mock stars}

We use mock stars to quantitatively assess the performance of the calibration approach. We simulated the relative radial velocity of the mock stars as the observation from one of the spectrographs. We produce 30 exposures in 7 nights for these mock stars, including 200 constant stars with the ground true velocity within $\pm$\,0.1\kms\ with the measurement uncertainty in 0$\sim$2\,\kms, 10 non-constant stars with the radial velocity changes of 2$\sim$10\,\kms, and 40 
binary stars with the amplitude of 3$\sim$15\,\kms\ in the period of 0.5$\sim$3 days. 

\begin{figure*}[!t]
  \centering
  \subfigure[]{
  \includegraphics[width = 6.5cm]{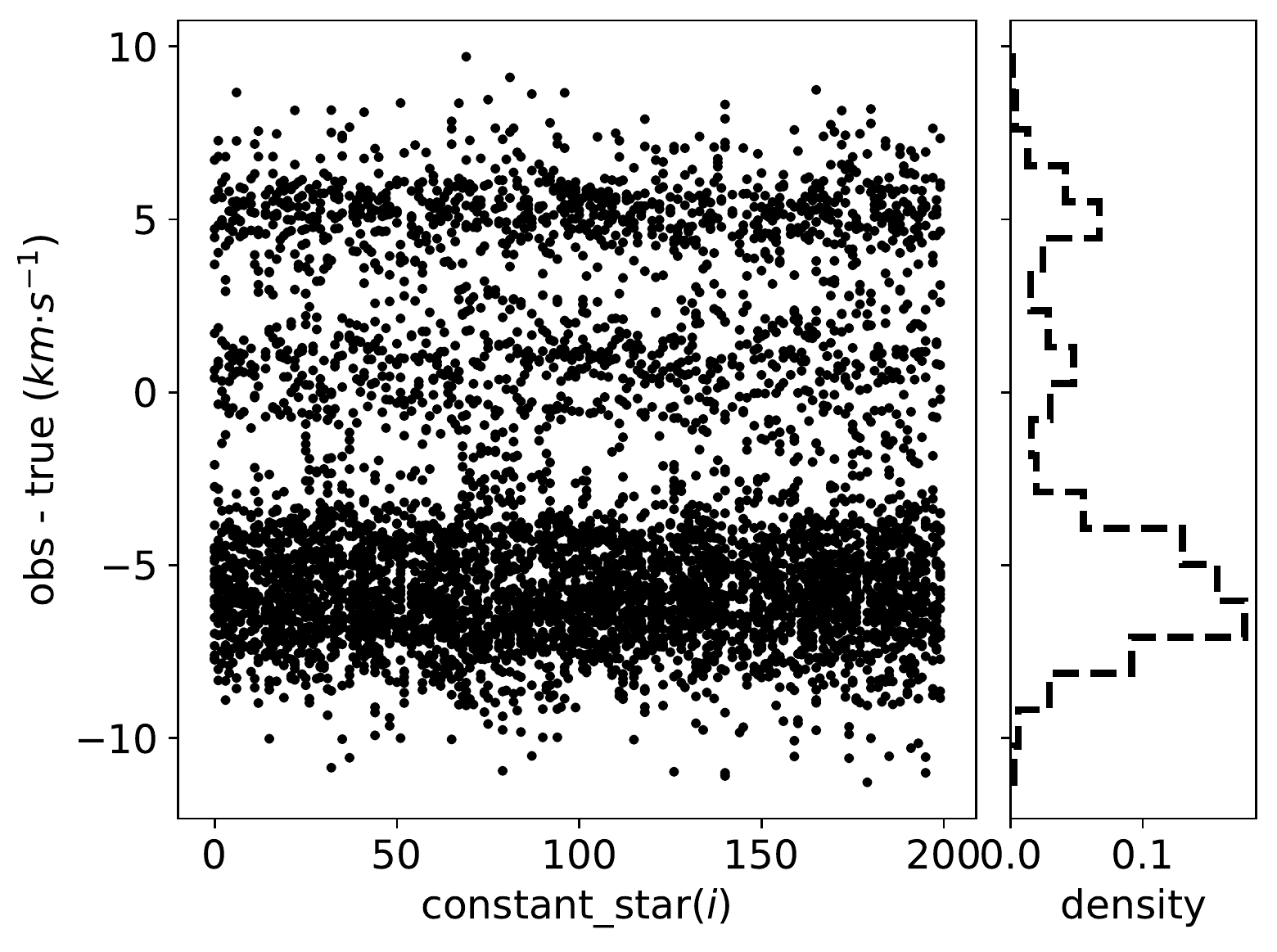}
}
  \subfigure[]{
  \includegraphics[width = 6.5cm]{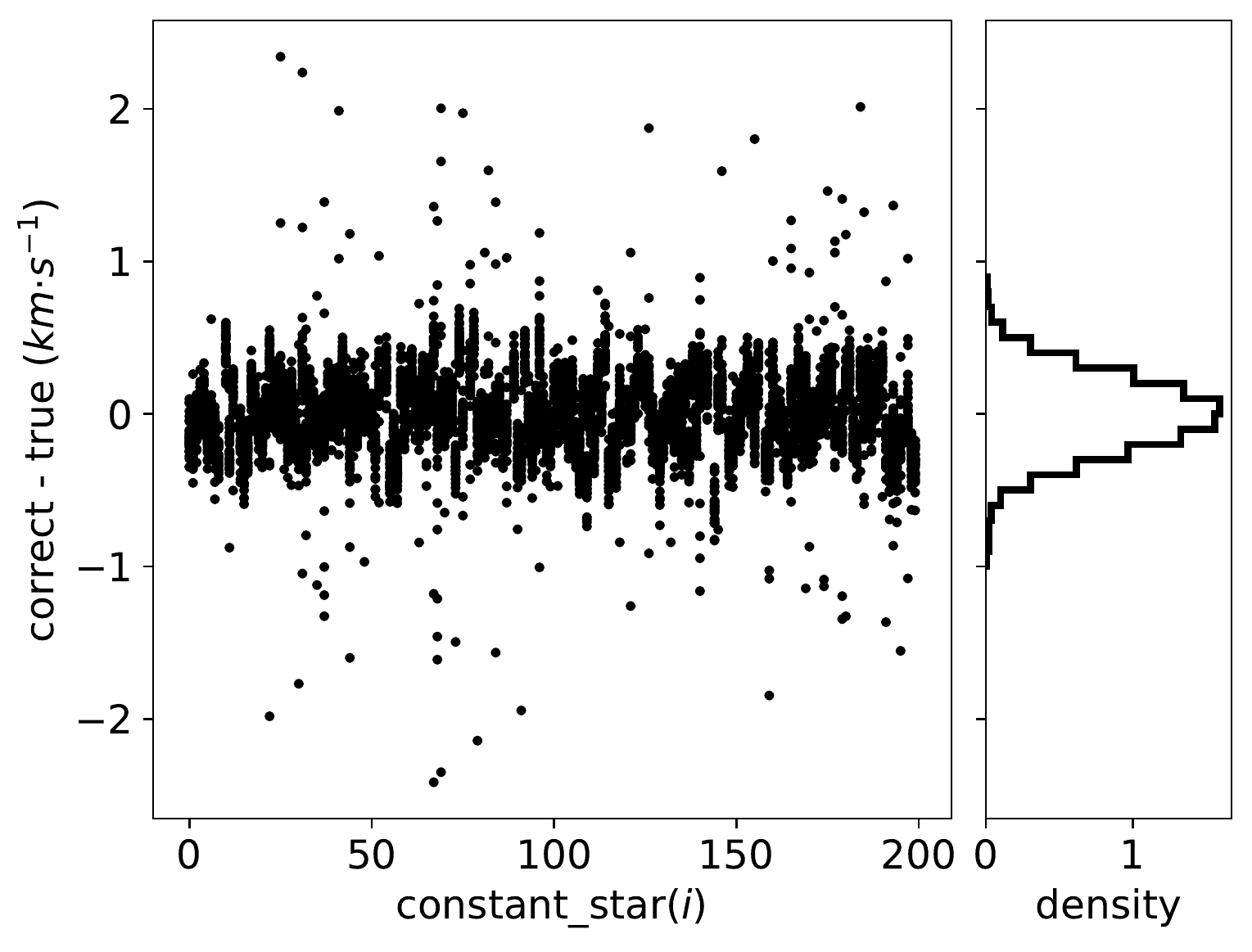}
} 
  \caption{The result of mock stars. Panel (a) shows the distribution of the difference between the directly measured radial velocities and the ground true velocities for mock constant stars. Panel (b) shows the residual value between the corrected velocities and the ground true velocities for mock constant stars.}
  \label{fig:1}
  
\end{figure*}

Then we add the systematic shift to these mock stars as the radial velocity that measured directly from the observation. The systematic shift of each night is randomly generated in the range of -7$\sim$7\kms. ~In the mean time, each exposure of the mock stars are added a random error with a Gaussian error of 0.5\,\kms.
After systematically shifting the radial velocities of the mock stars from zero, the constant stars' RV distribution becomes larger, and the difference between the observation and the ground true velocity is shown in figure~\ref{fig:1} (a). The residual values between the corrected radial velocities and the ground true velocities are shown in panel (b). It is seen that the mean value of the residual velocities is essentially between $\pm$0.50\,\kms, except non-constant stars.

\begin{figure*}[!t]
  \centering
  \subfigure[]{
    \includegraphics[height=4cm, width=4.5cm]{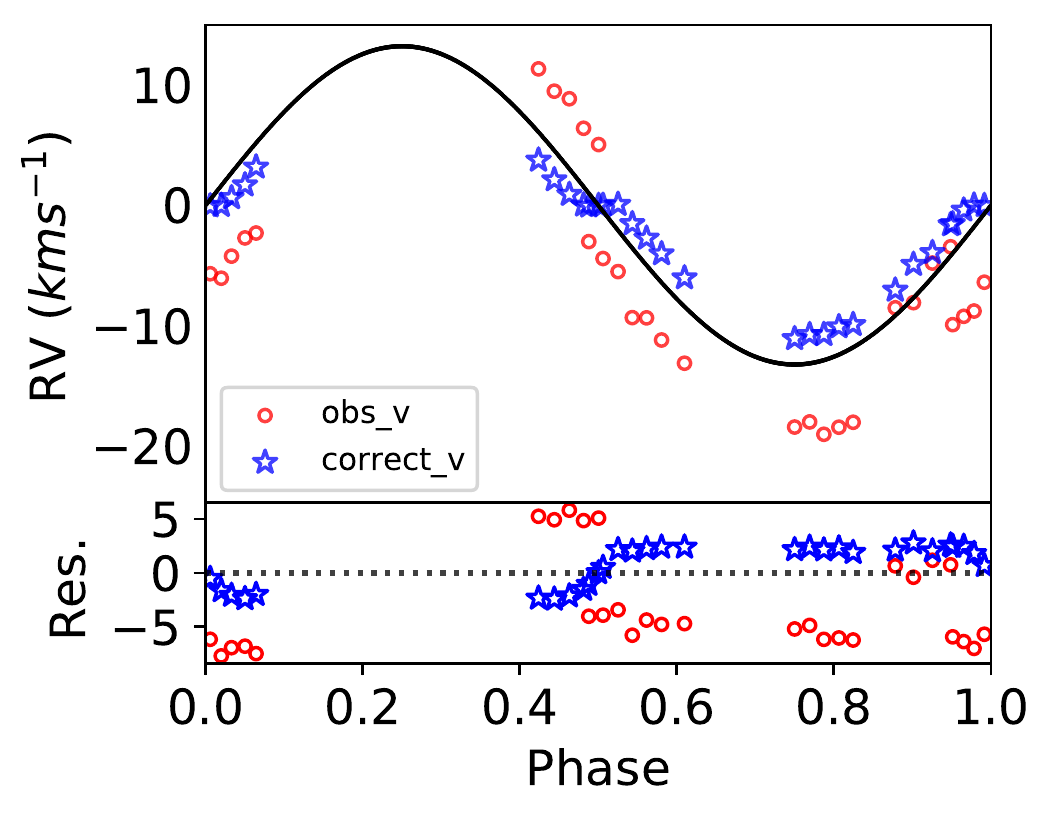}
    }
  \quad
  \subfigure[]{
    \includegraphics[height=4cm, width=4.5cm]{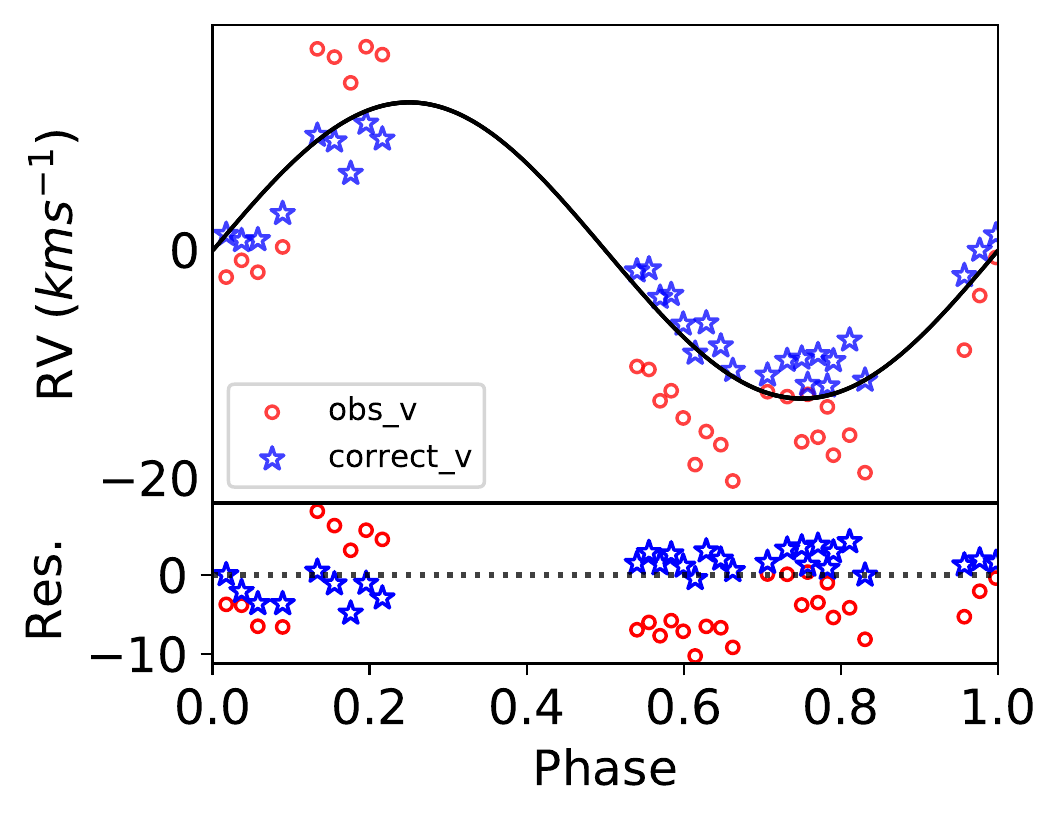}
    }
  \quad
  \subfigure[]{
    \includegraphics[height=4cm,width=4.5cm]{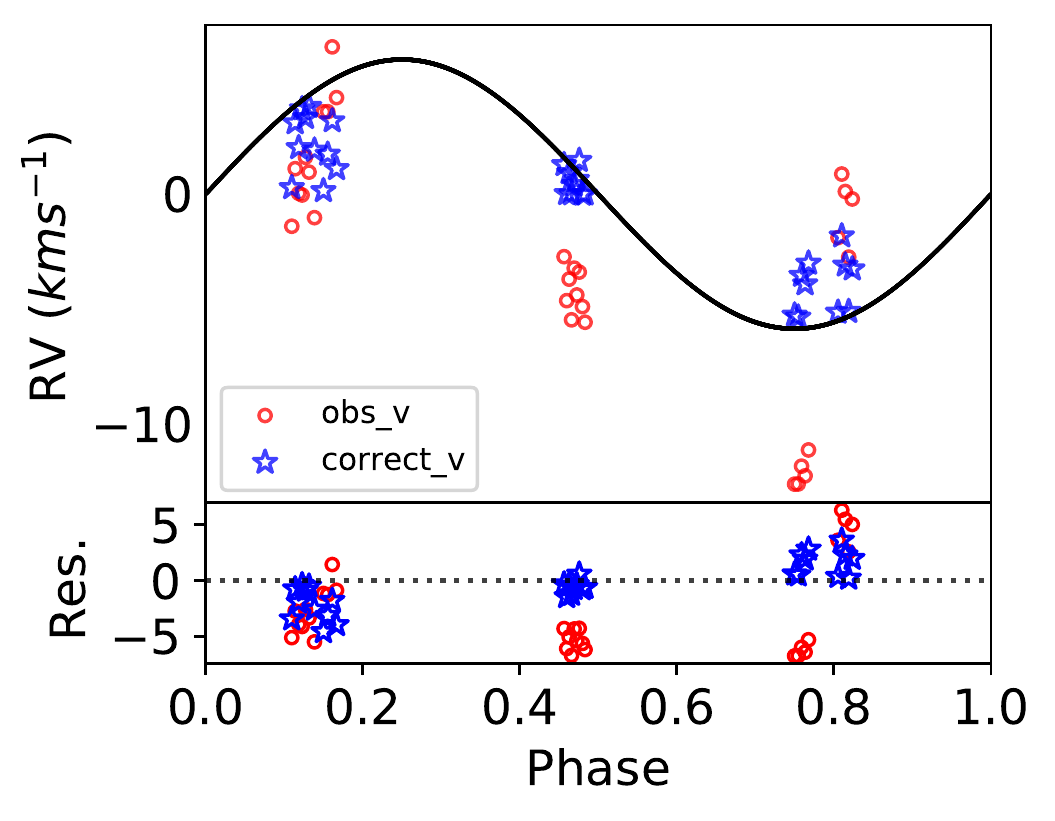}
    }
  
  \caption{The velocity correction for three mock periodic variable stars with different velocity amplitudes. The black-solid line indicate the true radial velocity curve. The red circles show the directly measured radial velocities with shifts. The blue stars display the corrected radial velocities. The bottom panels display the residual values with ground true velocity of non-corrected and corrected velocities with red circles and blue stars, respectively. 
  }
  \label{fig:2}
\end{figure*}

In Figure~\ref{fig:2}, we show the velocity curves of three mock binary stars. The top panels in figure~\ref{fig:2} show the velocity curves in black-solid lines, the mock observed radial velocities are shown in red circles, and the corrected velocities are shown in blue stars. The residual value is shown in the bottom panels of figure~\ref{fig:2}, including the residual values between the mock observational radial velocities and the true velocity curves in red circles, and the residual value between corrected velocities with the true velocity curves in blue stars. It is seen that the correction method can well reconstruct the radial velocity curves of the binary stars. After correction, the residual value are smaller than before correction. The non-costant offeset showing in the residual panel of the figure is due to that we have added a random RV measurement error to the simulation data (ranging from 0 to 2 \kms). After correction of the systematic offset in each observation, the random error of RV added to the mock data still exists, leading to the larger deviations than $\sim$ 0.5 \kms\, in the residuals. The detailed orbital parameters of the three mock binary stars are list in Table~\ref{tab:1}. 

\begin{table}
\caption{The detailed parameters of mock binaries shown in figure~\ref{fig:2} .}
\centering
\begin{tabular}{cccccc}
\hline
\hline
  \multicolumn{1}{c}{panel} &
  \multicolumn{1}{c}{Period} &
  \multicolumn{1}{c}{velocity amplitude} &
  \multicolumn{1}{c}{inclination} &
  \multicolumn{1}{c}{eccentricity} &\\
    & ($days$) & (\kms) & (\degree) & \\
\hline
(a) & 0.6894 & 13.23 & 88.65 & 0\\
\hline
(b) & 0.6373 & 13.01 & 88.15 & 0\\
\hline
(c) & 2.9098 & 5.91 & 81.79 & 0\\
\hline
\end{tabular}
\label{tab:1}
\end{table}

\section{Result} 
\label{sec:result}

 We apply the approach of calibration based on relative radial velocities (RV) to the LAMOST-MRS spectra with the SNR larger than 5 that released in DR7 time-domain plate, approximately 2,215,918 spectra, including 1,170,445 of the red arm and 1,045,473 of the blue arm. Then we determined the relative radial velocity in the red and blue arms separately. Each MRS spectrum is composed of two parts, the red covering from 6300\,\AA\ to 6800\,\AA\ and the blue covering from 4950\,\AA\ to 5350\,\AA. And for measuring the relative radial velocity, the wavelength from 6400\,\AA\ to 6700\,\AA\ of the red arm is used, and the wavelength from 5000\,\AA\ to 5300\,\AA\ from the blue arm is used. Then, the wavelength of each part of the spectrum is calibrated separately based on. The precision of wavelength calibration is better in the red than in blue, since the number of lamp spectra lines in the red part is more than that in the blue part. Meanwhile, the SNR of red spectra is in principle better than the blue arm due to larger throughput in red part. Therefore, we measure the relative radial velocity only using the red part spectra in the first step. Then we correct the relative radial velocities measured from the red part spectra. In the mean time, the candidate constant stars in each spectrograph are identified during the correction of radial velocities measured from red part spectra. Then, we directly use these candidate constant stars to correct the systematic shift induced in the wavelength calibration by the blue part spectra.

\begin{figure*}[!t]
\centering
\subfigure[]{
    \includegraphics[height=4cm,width=7.3cm]{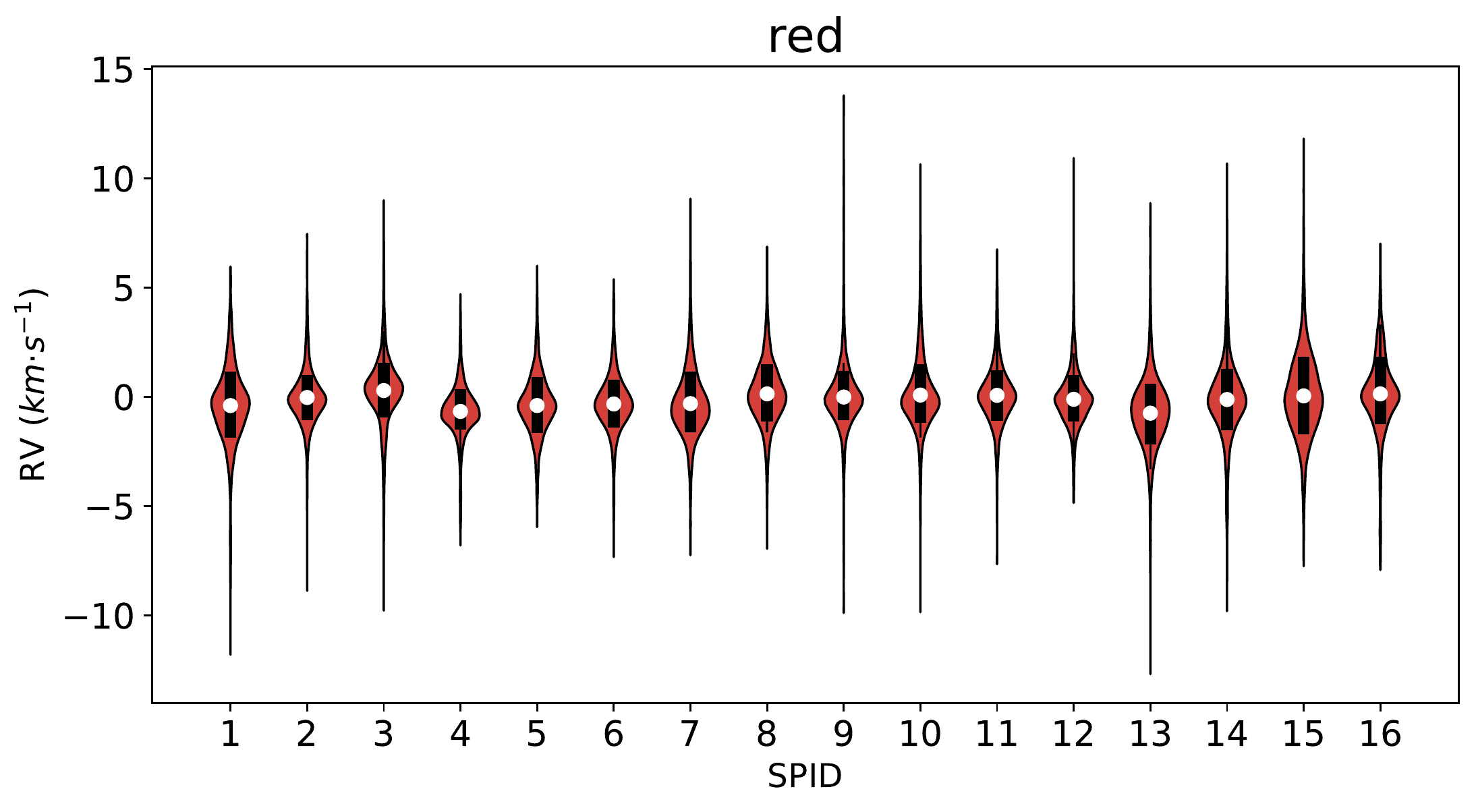}
    }
  \subfigure[]{
    \includegraphics[height=4cm,width=7.3cm]{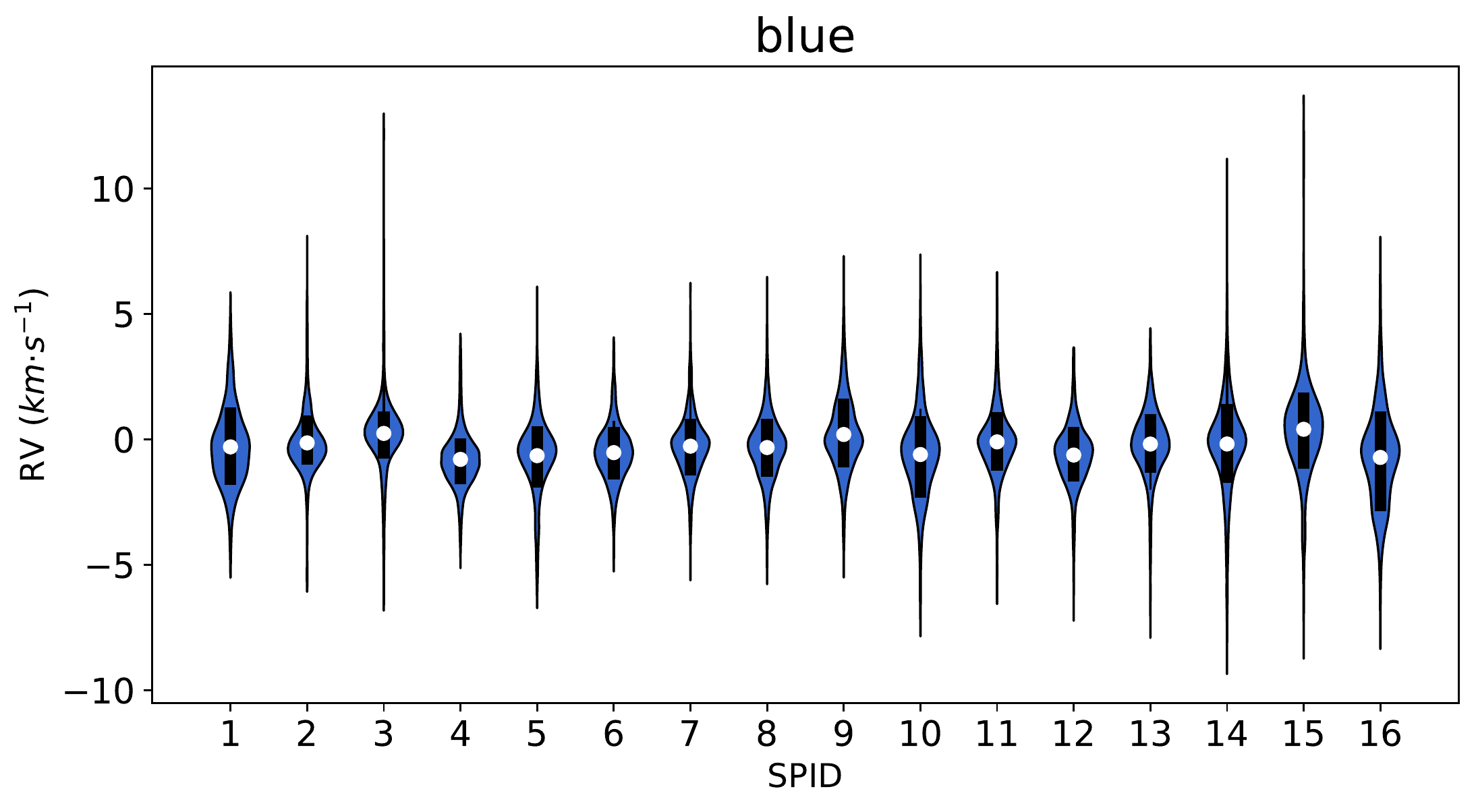}
    }
  \\
  \subfigure[]{
    \includegraphics[height=4cm,width=7.3cm]{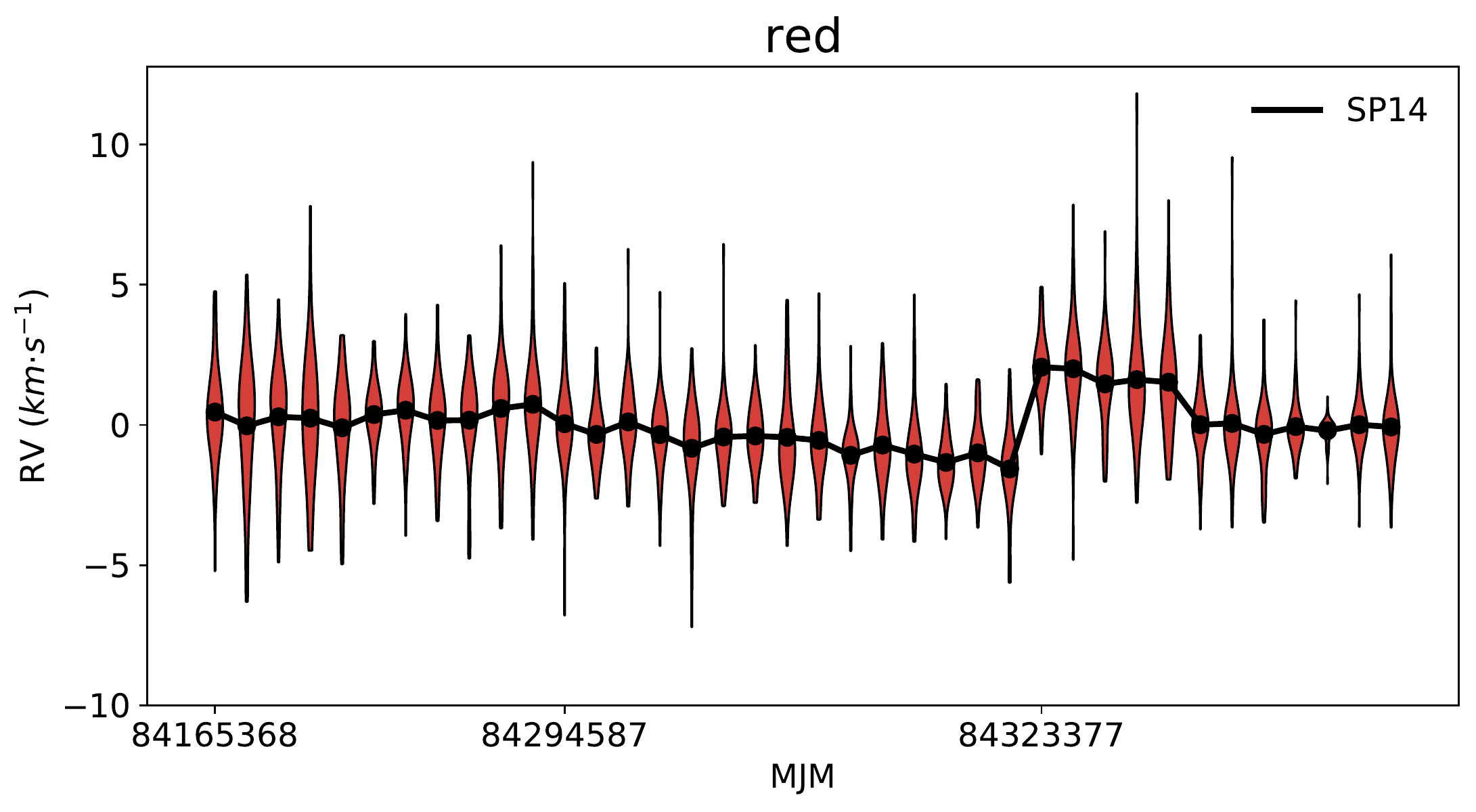}
    }
  \subfigure[]{
    \includegraphics[height=4cm,width=7.3cm]{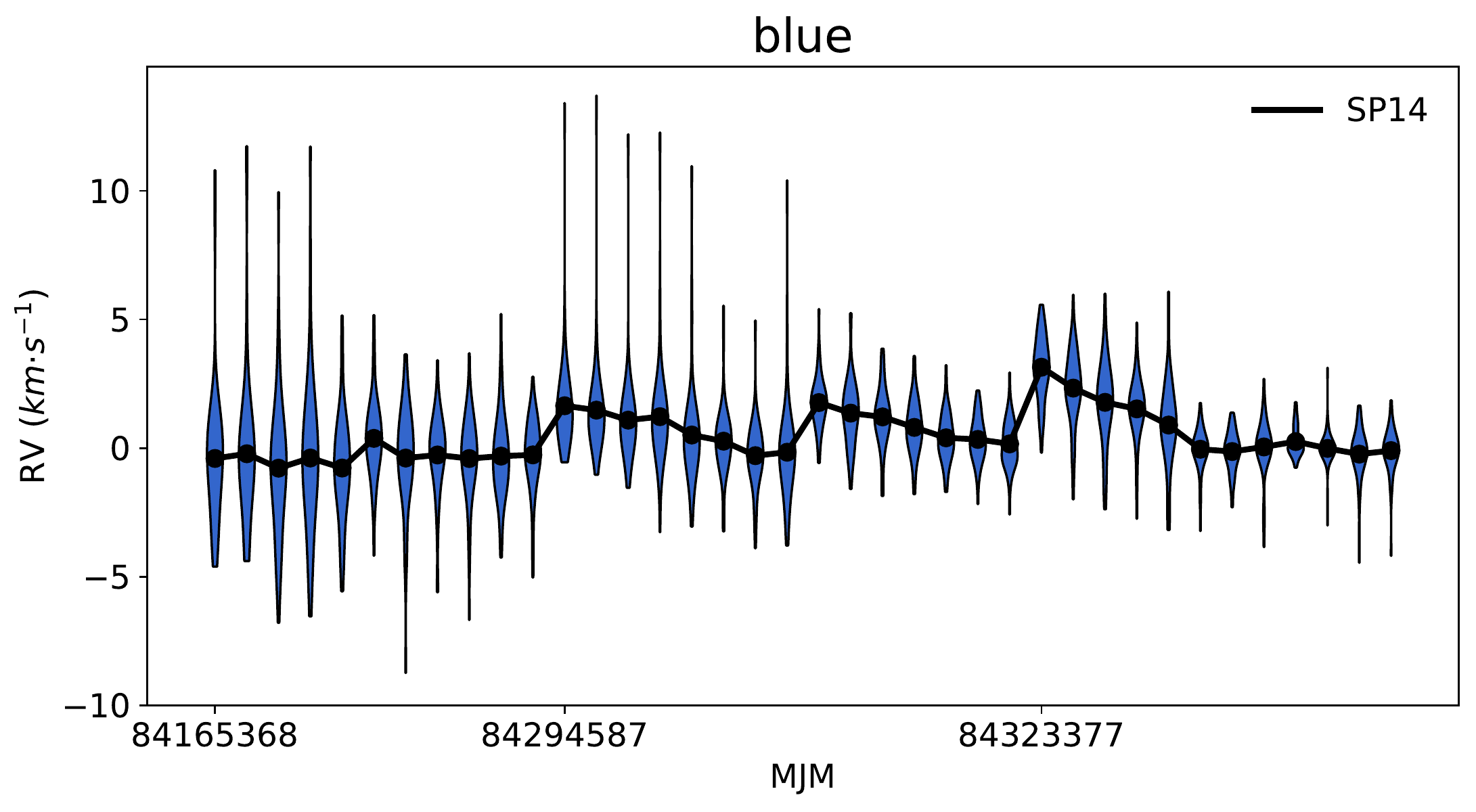}
    }
\caption{Panels (a) and (b) show the distribution of the directly measured relative radial velocity from the spectra of constant star candidates from one sample field of LAMOST MRS in 16 spectrographs of red and blue arms. The circles are the mean relative radial velocity of candidate constant stars. Panels (c) and (d) show the distribution of the directly measured relative radial velocity from the spectra observed by LAMOST MRS of constant star candidates for a spectrograph in red and blue arms, respectively, but in different exposures. The black dots indicate the initial mean radial velocities in different exposures.}
\label{fig:3}
\end{figure*}

Figure~\ref{fig:3} (a) and (b) show the distribution of the directly measured relative radial velocity from the spectra of constant star candidates from one sample field of LAMOST MRS in 16 spectrographs of red and blue arms, respectively. The violin-plot shows the dispersion of the relative radial velocities of constant star candidates in each spectrograph, and the white spots are the mean relative radial velocities of these constant star candidates in each spectrograph. It illustrates that the radial velocity zero-points are slightly different between all spectrographs. And figure~\ref{fig:3} (c) and (d) show the relative radial velocity distribution of constant star candidates for a spectrograph in 38 different exposures. The violin-plot shows the dispersion of the relative radial velocities of constant star candidates in each exposure, and the black dots are the mean relative radial velocities of these constant star candidates. It can be seen that the mean velocity of the candidate constant stars can move to as much as 3 \kms\ from zero in some observations, especially in different nights. And according to panels (c) and (d), the variation of the mean radial velocities of red arm is more stable than those in blue arm. To sum up, we perform calibration for each spectrograph separately, and apply the calibration on red arm at first.

\begin{figure*}[!t]
  \centering
  \subfigure[]{
    \includegraphics[height=6cm,width=7cm]{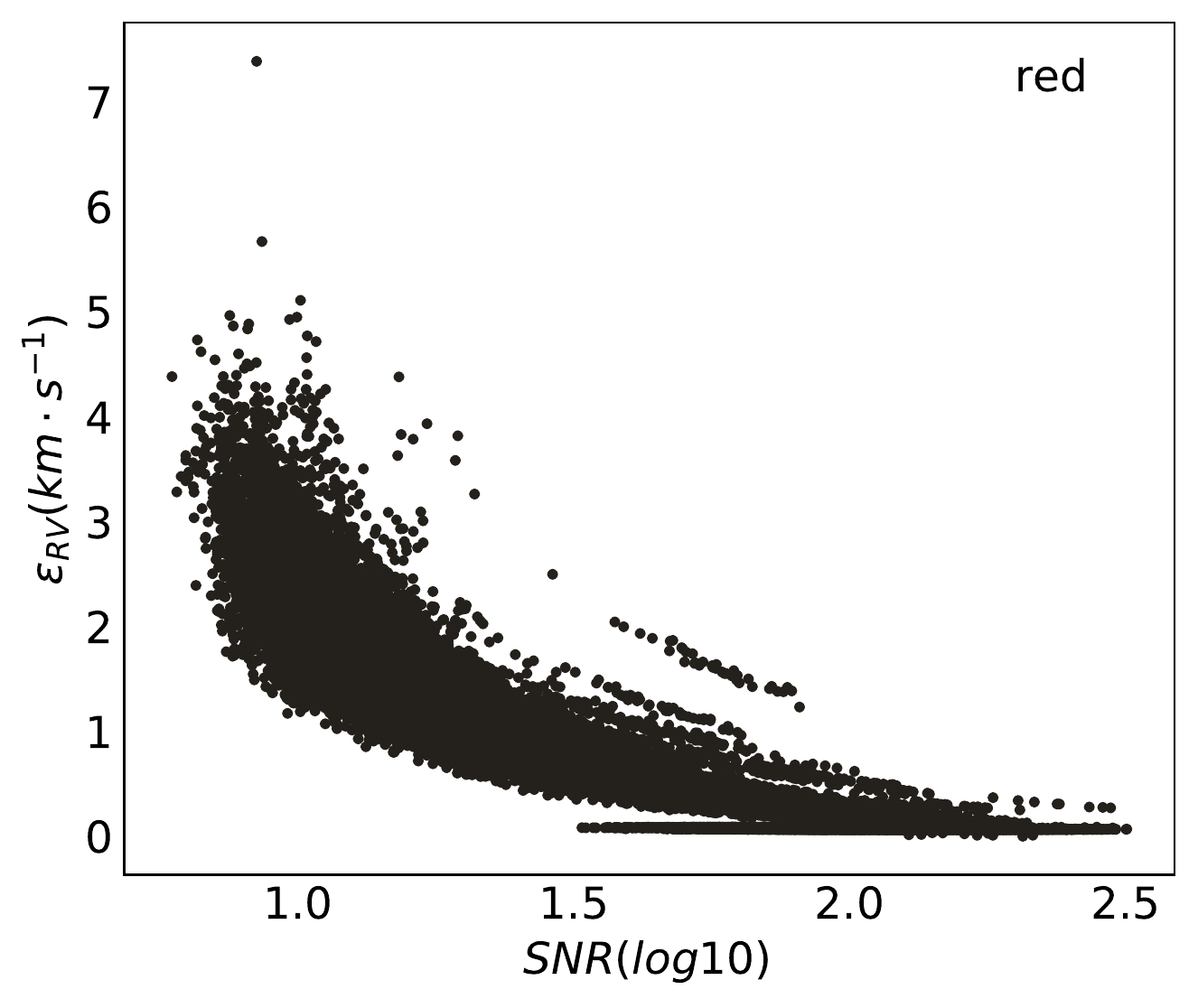}
    }
  \subfigure[]{
    \includegraphics[height=6cm,width=7cm]{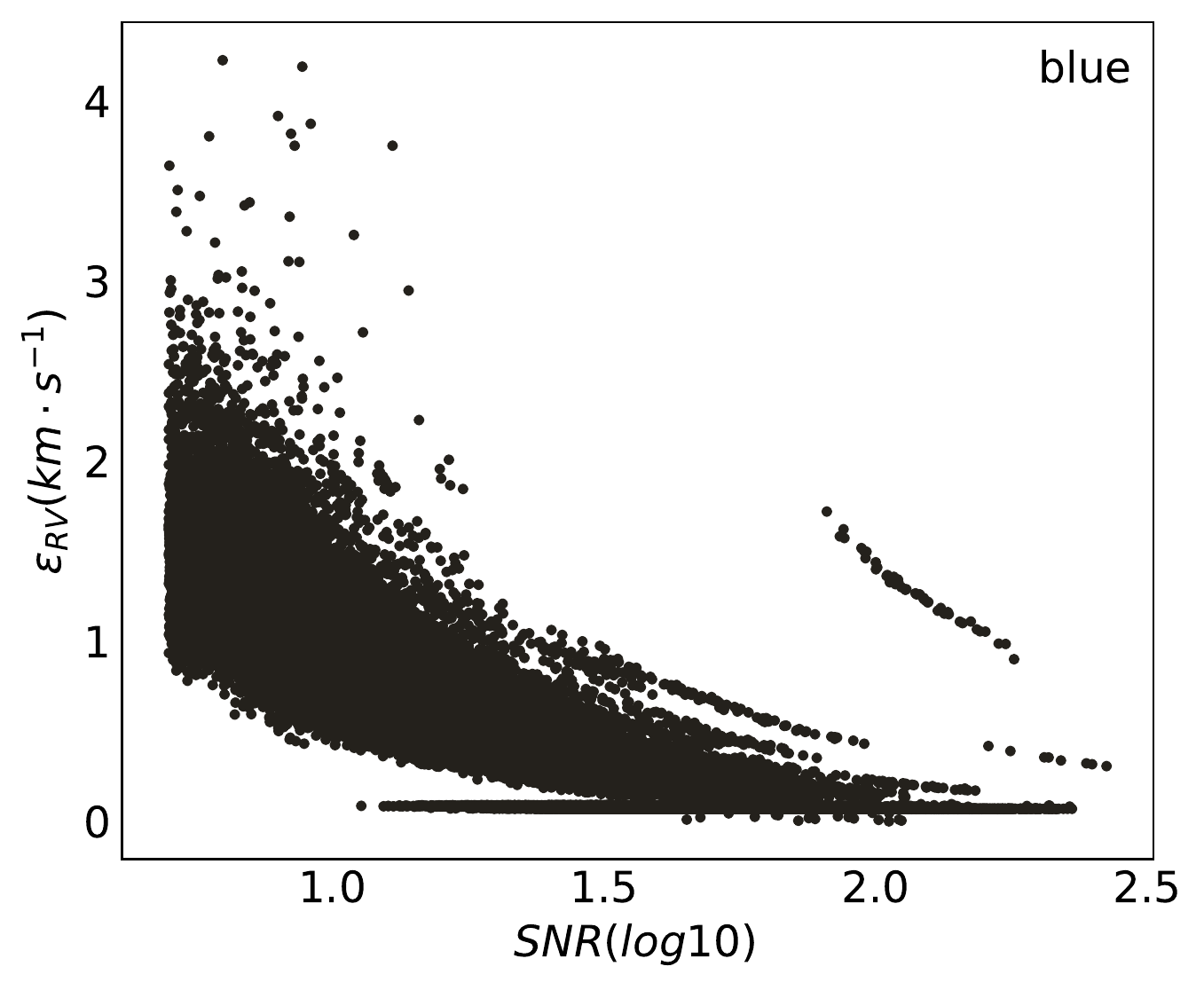}
    }

  \caption{Panels (a) and (b) show the relation between measurement uncertainty ($\epsilon_{RV}$) with SNR that selected from the same sample field with figure~\ref{fig:3} of LAMOST MRS in the red and blue arm, respectively.}
  \label{fig:4}
  
\end{figure*}

Figure~\ref{fig:4} (a) and (b) show the relation between signal-to-noise ratio(SNR) with the measurement uncertainty of radial velocity from a same sample field same as figure~\ref{fig:3} of LAMOST MRS in red and blue arm, respectively, and the measurement uncertainty of radial velocity is calculated as the standard deviation of the best-fit Gaussian distribution corresponding to the likelihood function distribution, and donated as $\epsilon_{RV}$. Panels (a) and (b) both show the decreasing uncertainty of measurement with the increasing SNR. The non-physical strips are found in fig~\ref{fig:4}, it is caused by the multi-star system such as binaries. For these stars, the regular RV measurement method can not identify the radial velocity of their components correctly, resulting in a large measurement uncertainty for some spectra in high SNR. And it can be seen that the uncertainty of measuring relative radial velocity of red arm is relatively larger than the blue arm due to the fact that $H_{\alpha}$ is usually the only prominent feature in the red arm, although it is more stable than blue arm. So we calibrate the systematic shift for red arm at first, and the constant star candidates picked out after the red arm calibration are directly used in the blue arm. Using these constant star candidates, the systematic shift of different exposures for blue arm is then be corrected.

\begin{figure*}[!t]
  \centering
  \subfigure[]{
    \includegraphics[height=6cm,width=7cm]{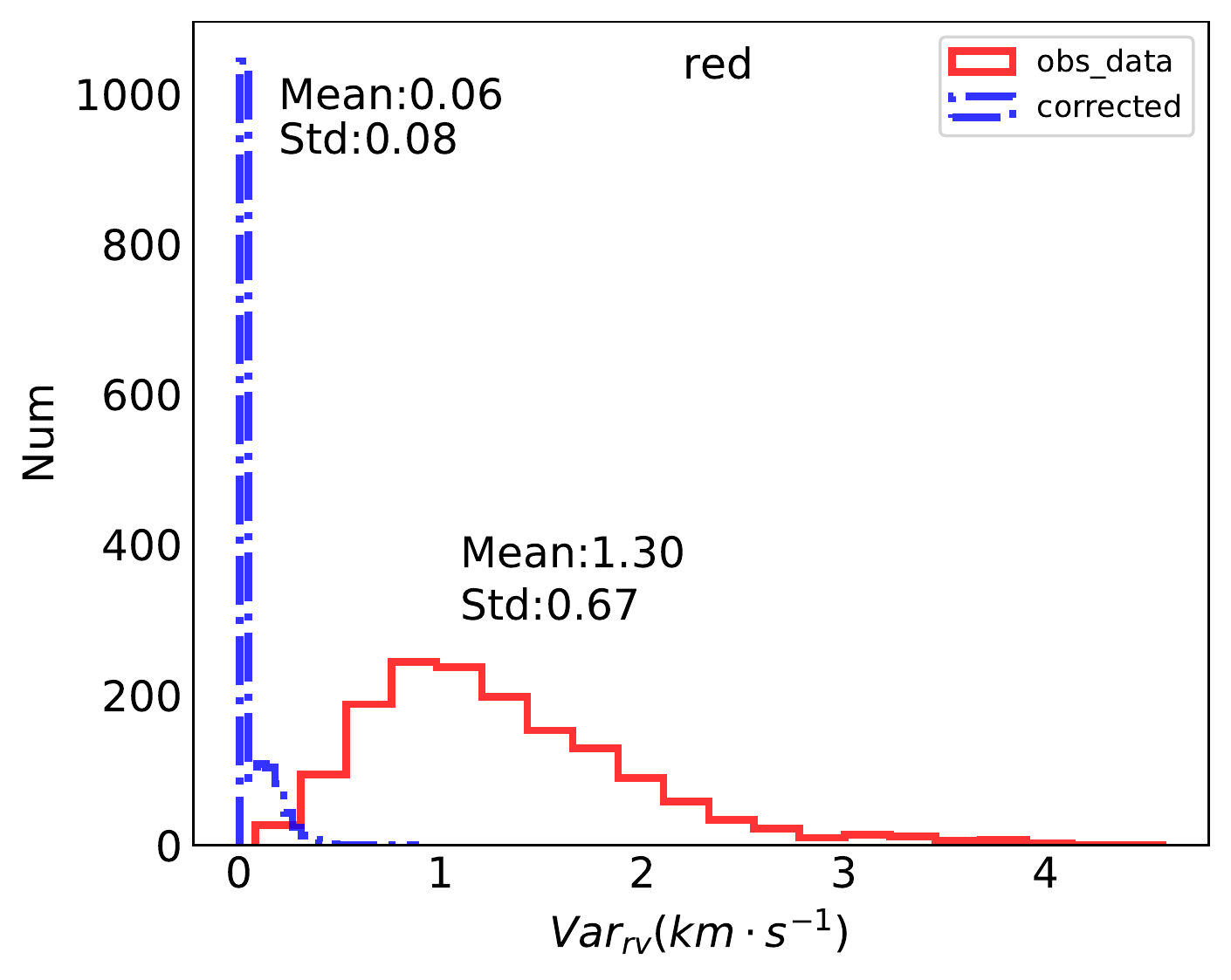}
    }
  \quad
  \subfigure[]{
    \includegraphics[height=6cm,width=7cm]{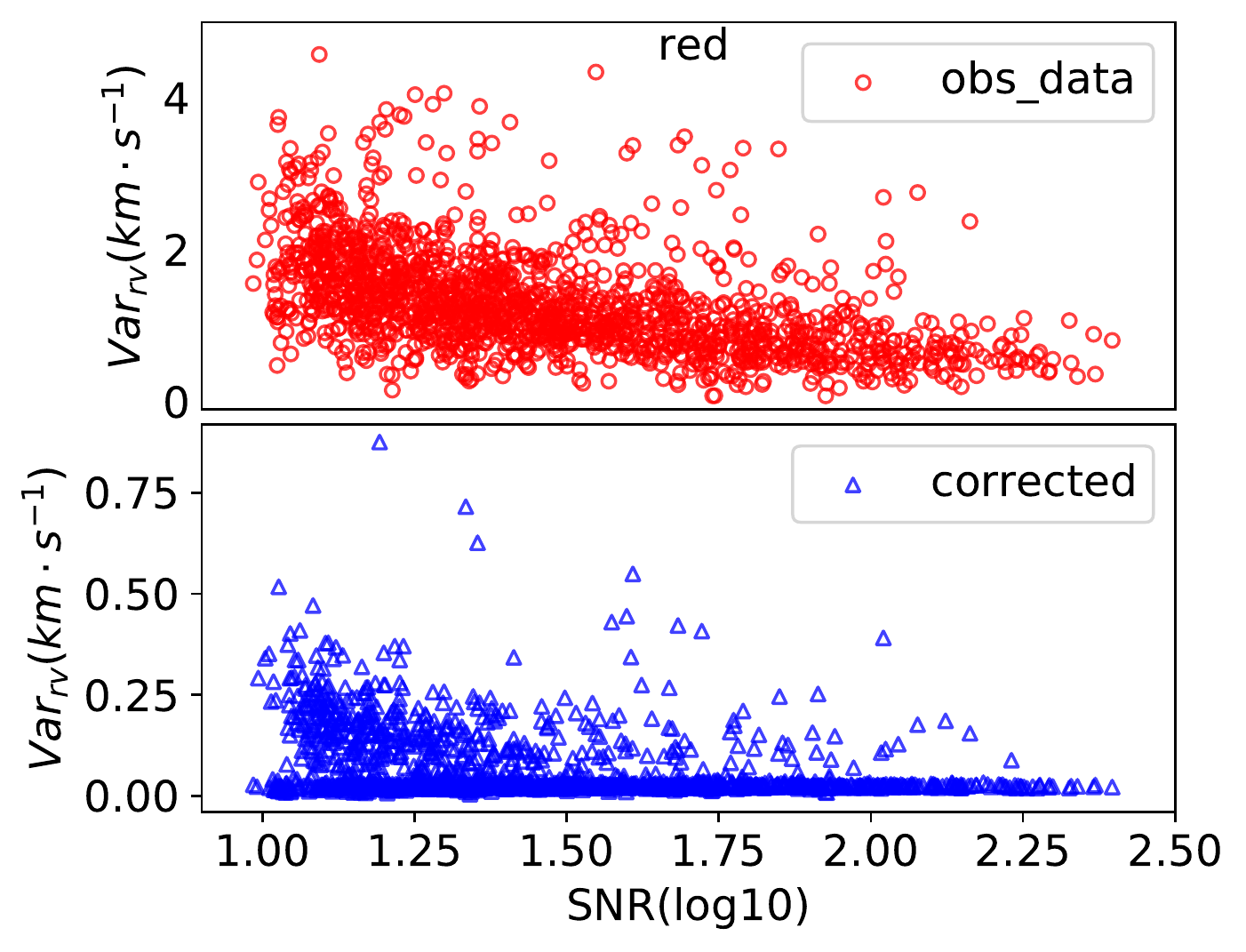}
    }
  \subfigure[]{
    \includegraphics[height=6cm,width=7cm]{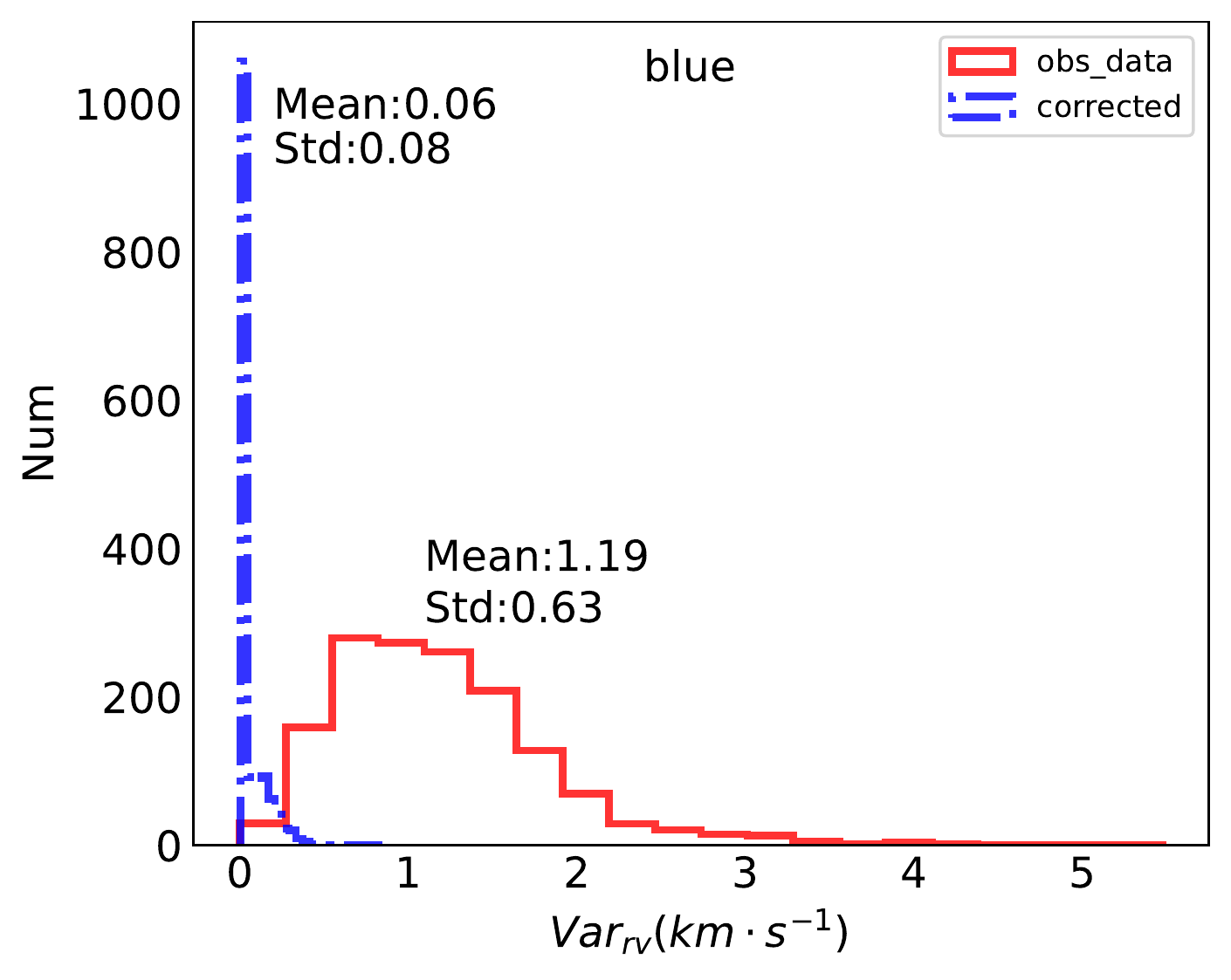}
    }
  \quad
  \subfigure[]{
    \includegraphics[height=6cm,width=7cm]{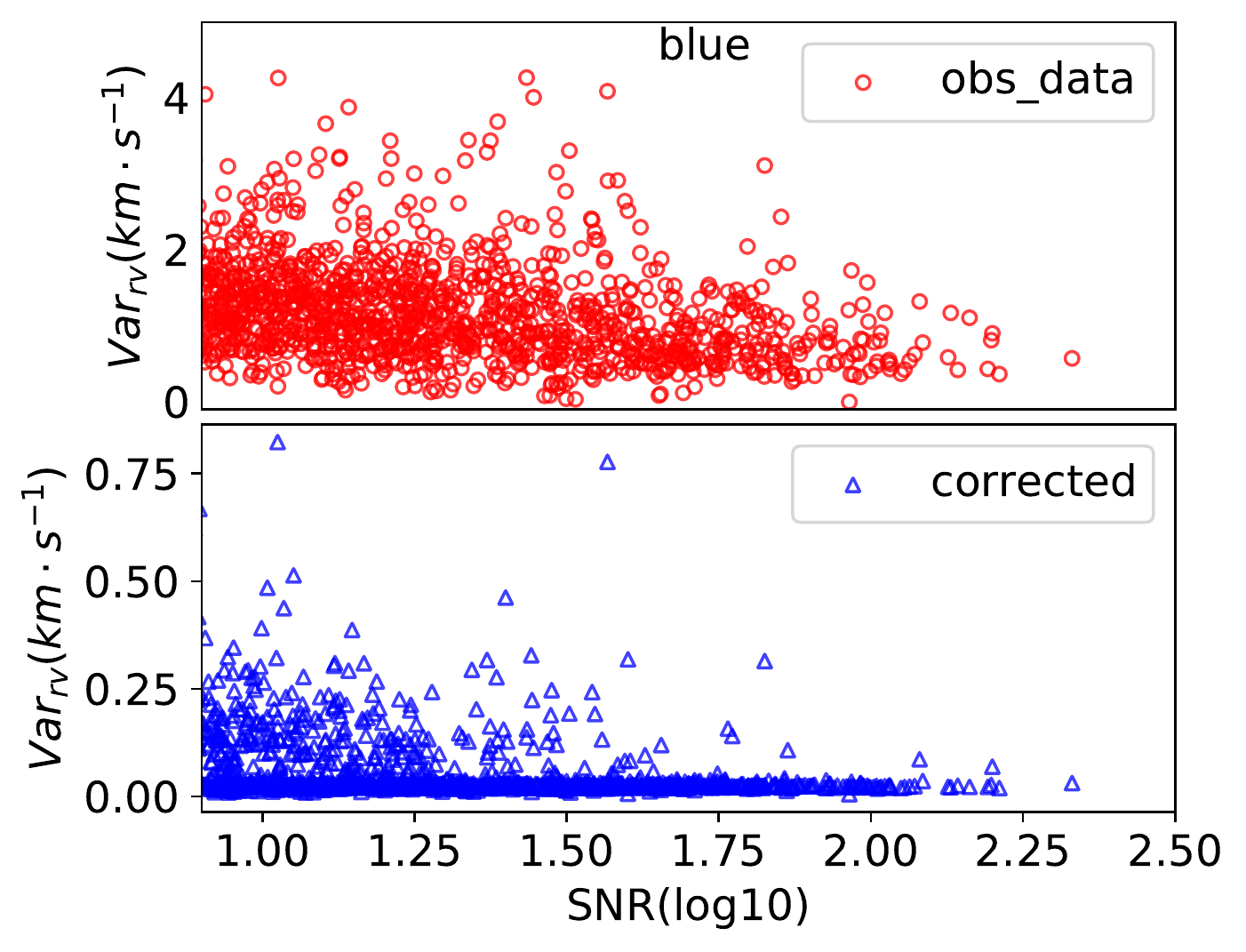}
    }
  \quad 
  \caption{The non-calibration and calibrated radial velocity variation ($Var_{rv}$) from the same sample field with figure~\ref{fig:3} of LAMOST MRS for selected candidate constant stars, and their relation with SNR for two arms. In panels (a) and (c) the red solid, blue dash-dotted histograms show the $Var_{rv}$ distributions for constant stars with non-calibration and calibrated in the red and blue arm, respectively. In panel (b) and (d), the red circles, blue triangles show the distributions of $Var_{rv}$ for constant stars with non-calibration and calibrated in red and blue arm, respectively.}
  \label{fig:5}
\end{figure*}

 Figure~\ref{fig:5} shows the distribution of the non-calibrated and calibrated radial velocity variation in for candidate constant stars from the same sample field as figure~\ref{fig:3}. The radial velocity variation is measured by the standard deviation of the radial velocities of an constant candidates over different exposures, donated as $Var_{rv}$. In the mean time, the relations between the radial velocity variation of the two arms and SNR are also shown in figure ~\ref{fig:5}. Panel (a) shows the non-calibrated constant candidates' $Var_{rv}$ as red solid histogram, which are directly measured from observational spectra in the red arm, represents the systematic shift of different exposures, the mean value of the non-calibrated variation is around 1.30\,\kms.~And the calibrated $Var_{rv}$ of constant stars candidates is shown as blue dash-dotted histogram, which tightly concentrates to around zero. Panel (b) shows the relation of $Var_{rv}$ distribution and SNR of constant candidates, with the top panel showing the $Var_{rv}$ distribution for direct measurement from the observational spectra as red circles and the bottom show the $Var_{rv}$ distribution of constant candidates after calibration as blue triangles. After calibrating the systematic shift, the $Var_{rv}$ of constant stars are almost below 0.5\,\kms.~As panel (b) shows, the calibration precision depends on the SNR, and the relatively lower precision results in high SNR is because of the relatively large variation of SNR for the same target in time-domain observation. Panels (c) and (d) show similar distributions for blue-arm calibration result with panels (a) and (b), respectively. In the blue arm, the initial constant candidates are adopted from red-arm correction result. Panel (c) shows the non-calibrated $Var_{rv}$ of blue-arm as red histogram with the mean of 1.19\,\kms,~and the calibrated $Var_{rv}$ is shown in blue dash-dotted histogram, which has mean $Var_{rv}$ of 0.06\,\kms. ~Panel(d) also shows the decreasing $Var_{rv}$ of constant stars with increasing SNR, and after calibrating, the $Var_{rv}$ is almost below to 0.5\,\kms.

In total, 2,215,918 spectra from the 59 time-domain plates of LAMOST MRS of DR7 were calibrated in this work, including 1,170,445 of the red arm and 1,045,473 of the blue arm. The catalog of calibrated relative radial velocity for these time-domain observations is listed in Table~\ref{tab:2}. which including ra, dec, hjd, SNR, directly measured relative radial velocity (DRV\_r and DRV\_b), the uncertainty of measurement (DRV\_r\_err and DRV\_b\_err), and the calibrated result (DRV\_r\_corr and DRV\_b\_corr), and the ``r'' and ``b'' label the red and blue arm, respectively. The whole catalog is available in its entirety in machine-readable form.

After self-calibration in red and blue arms, we obtain 65,123 constant star candidates. And the catalog of these constant star candidates is also available in machine-readable form. As the observations increasing, these constant samples will be evolved and become more accurate. This is a long-term process that can be used for LAMOST time-domain medium resolution survey.
\begin{table*}[h!]
\label{tab:2}
\scriptsize
\centering
\caption{The catalog of relative radial velocity for LAMOST MRS time-domain plates.}
% \resizebox{\textwidth}{!}{
\begin{tabular}{cc|ccccccccc}
\hline
\hline
  \multicolumn{1}{c}{ra} &
  \multicolumn{1}{c}{dec} &
  \multicolumn{1}{c}{hjd} &
  \multicolumn{1}{c}{SNR\_r} &
  \multicolumn{1}{c}{DRV\_r} &
  \multicolumn{1}{c}{DRV\_r\_err} &
  \multicolumn{1}{c}{DRV\_r\_corr} &
  \multicolumn{1}{c}{SNR\_b} &
  \multicolumn{1}{c}{DRV\_b} &
  \multicolumn{1}{c}{DRV\_b\_err} &
  \multicolumn{1}{c}{DRV\_b\_corr} \\
\hline
& & 2458411.145602 & 24.34 & -4.25 & 1.05 & 0.04 & 15.32 & -4.90 & 0.72 & 0.02\\
&& 2458416.177853 & 5.84 & -4.83 & 2.88 & 0.00 & 5.02 & -6.12 & 1.66 & 0.00\\
16.601013 & 3.819259 & 2458420.162413 & 15.43 & 0.51 & 1.56 & 0.03 & 9.21 & -0.26 & 1.06 & 0.00\\
  & & 2458449.087430 & 6.01 & -8.11 & 3.83 & 0.14 & 5.0 & -3.14 & 2.01 & 0.05\\
  && $\vdots$ & $\vdots$& $\vdots$ & $\vdots$ & $\vdots$ & $\vdots$ & $\vdots$ & $\vdots$ & $\vdots$\\
\hline
&& 2458411.145607 & 7.95 & -4.40 & 3.21 & 0.05 & 6.06 & -7.05 & 1.83 & -0.83\\
&& 2458439.058429 & 8.14 & 4.29 & 3.7 & 0.02 & 6.99 & -0.51 & 1.73 & 0.01\\
16.602702 & 4.362617 & 2458451.078222 & 11.4 & -1.95 & 2.86 & -0.03 & 10.9 & 0.06 & 1.24 & 0.01\\
&& 2458453.050072 & 13.61 & 0.73 & 2.33 & 0.02 & 10.35 & 0.27 & 1.29 & -0.02\\
&&$\vdots$ & $\vdots$ & $\vdots$& $\vdots$& $\vdots$ &$\vdots$ & $\vdots$ & $\vdots$&$\vdots$\\
\hline
&& 2458411.129353 & 42.14 & -3.37 & 0.51 & -0.02 & 21.26 & -3.77 & 0.37 & -0.01\\
&& 2458420.112249 & 15.56 & 0.35 & 1.20 & -0.05 & 7.45 & -1.22 & 0.81 & -0.03\\
16.604269 & 5.121394 & 2458439.058448 & 22.54 & 0.95& 0.83 & -0.03 & 11.16 & 0.83 & 0.58& -0.01\\
&& 2458451.061743 & 32.21 & -0.03 & 0.62 & 0.00 & 18.48 & -0.06 & 0.41 & -0.01\\
&&$\vdots$ & $\vdots$ & $\vdots$& $\vdots$& $\vdots$ &$\vdots$ & $\vdots$ & $\vdots$&$\vdots$\\
\hline
&& 2458411.145590 & 16.02 & -6.16 & 1.57 & 0.05 & 5.82 & -11.03 & 1.28 & -0.83\\
16.612777 & 2.792158 & 2458439.058398 & 12.91 & -0.19 & 1.83& 0.00 & 5.48 & -3.09 & 1.28 & 0.00\\
&&$\vdots$ & $\vdots$ & $\vdots$& $\vdots$& $\vdots$ &$\vdots$ & $\vdots$ & $\vdots$&$\vdots$\\
\hline
$\cdots$ & $\cdots$ &  $\cdots$ & $\cdots$ & $\cdots$& $\cdots$& $\cdots$ &$\cdots$ & $\cdots$ & $\cdots$& $\cdots$ \\
\hline
\hline
\end{tabular}
\end{table*}

\begin{figure*}[!t]
  \centering
    \includegraphics[height=16cm,width=16cm]{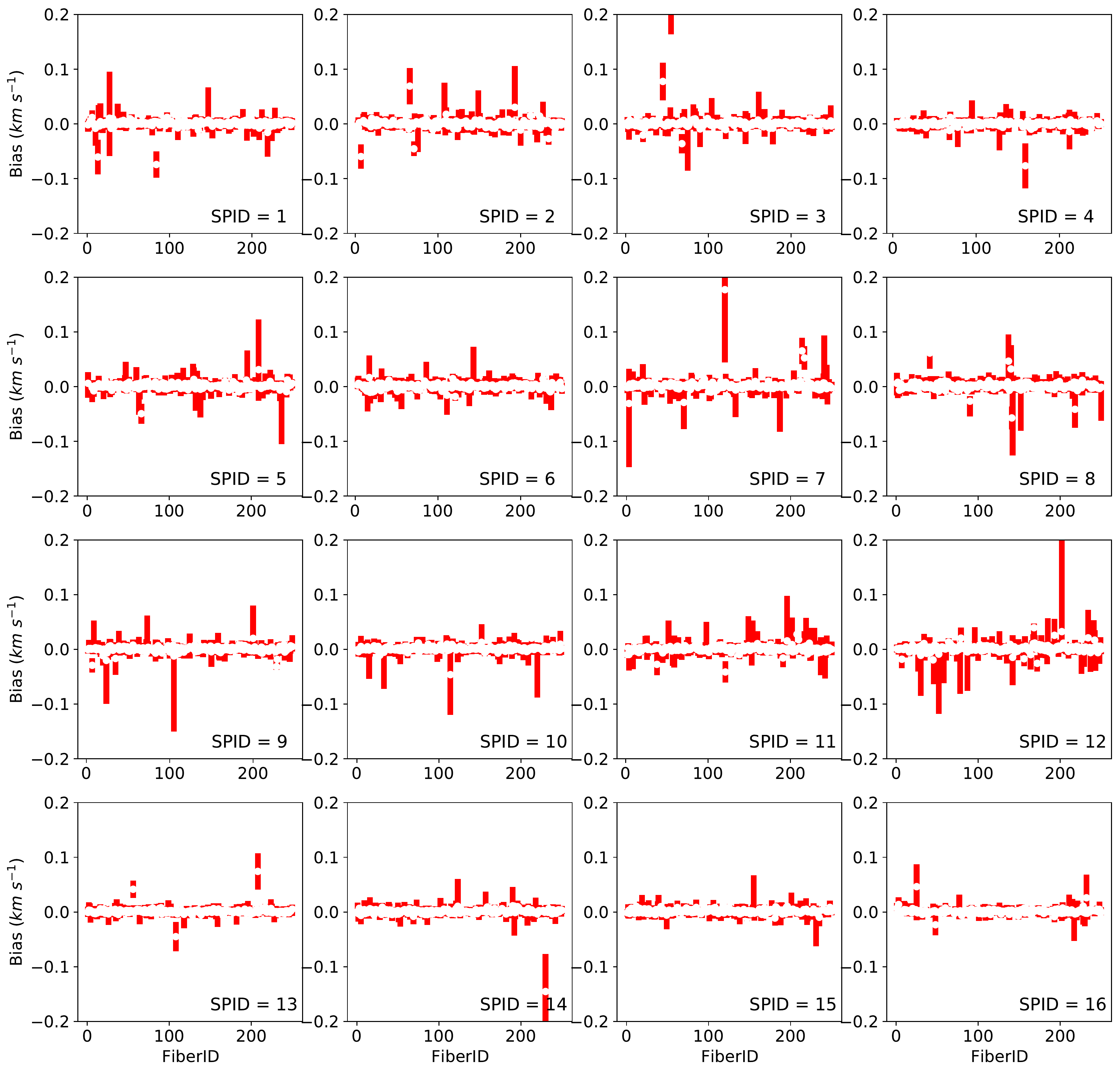}
  \caption{The RV bias due to fiber-to-fiber variations after calibration in red arm.}
  \label{fig:6}
  \end{figure*}

\begin{figure*}[!t]
  \centering
 \includegraphics[height=16cm,width=16cm]{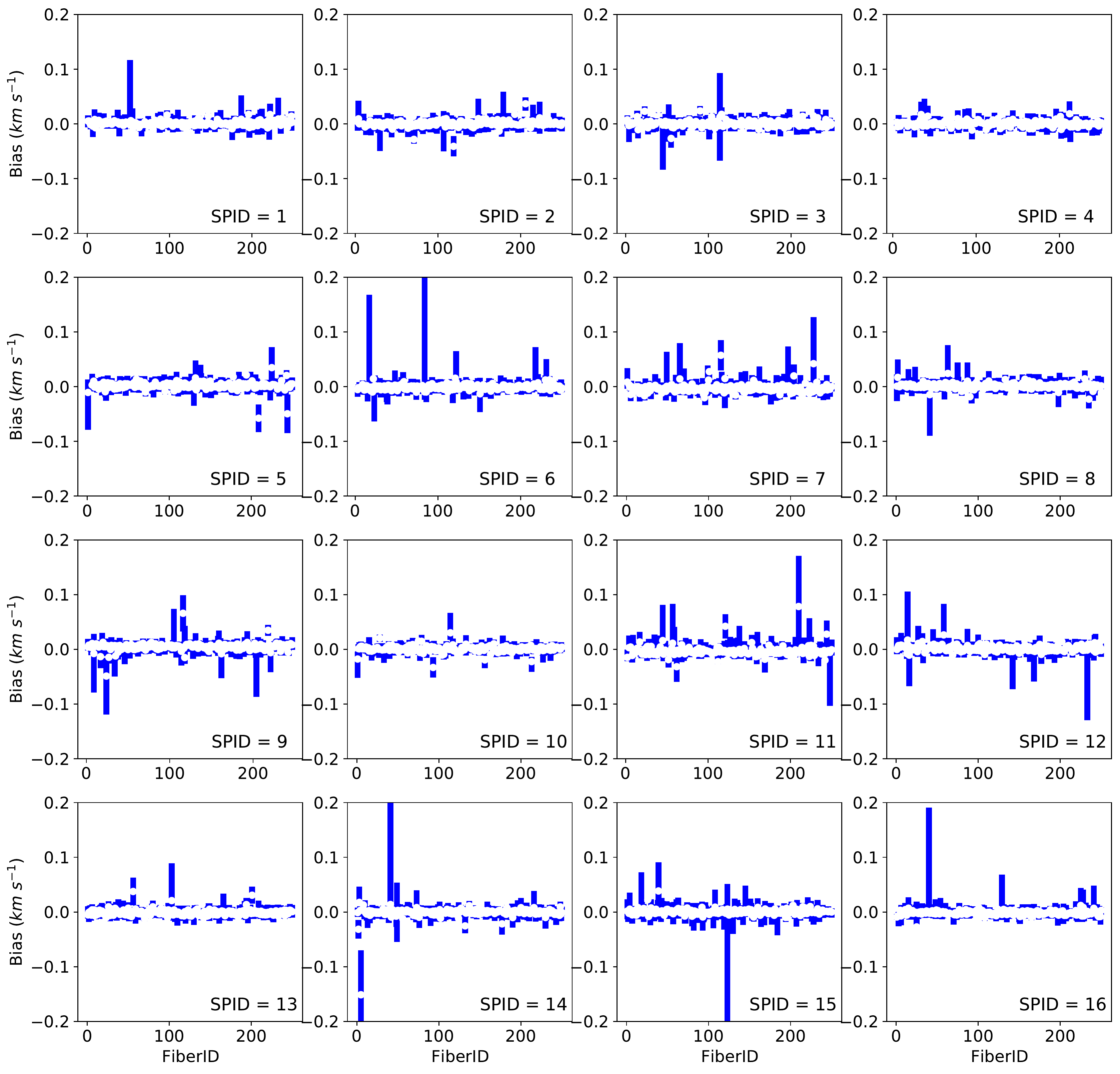}
  \caption{The RV bias due to fiber-to-fiber variations after calibration in blue arm.}
  \label{fig:7}
  \end{figure*}

Figure~\ref{fig:6} and figure~\ref{fig:7} show the typical RV variations for each fiber after calibration in red and blue arms, respectively. The vertical red and blue lines indicate the range of the RV variations of constant stars for each fiber. It indicates that the mean value of RV variations due to the fiber-to-fiber is around 0 with a dispersion of 0.1\,\kms\ , which is corresponding to the limit of the calibration.
  
\section{Discussion} \label{sec:Discussion}
The precision of the calibration is highly correlated with the precision of the radial velocity measurement method, and the measurement precision is limited by the spectra type and SNR. So, the precision of this relative radial velocity measurement method is discussed in different types of spectra, as well as the pulsating variable stars and binary system. 

\begin{figure*}[!t]
  \centering
  \subfigure[]{
    \includegraphics[height=6cm,width=7cm]{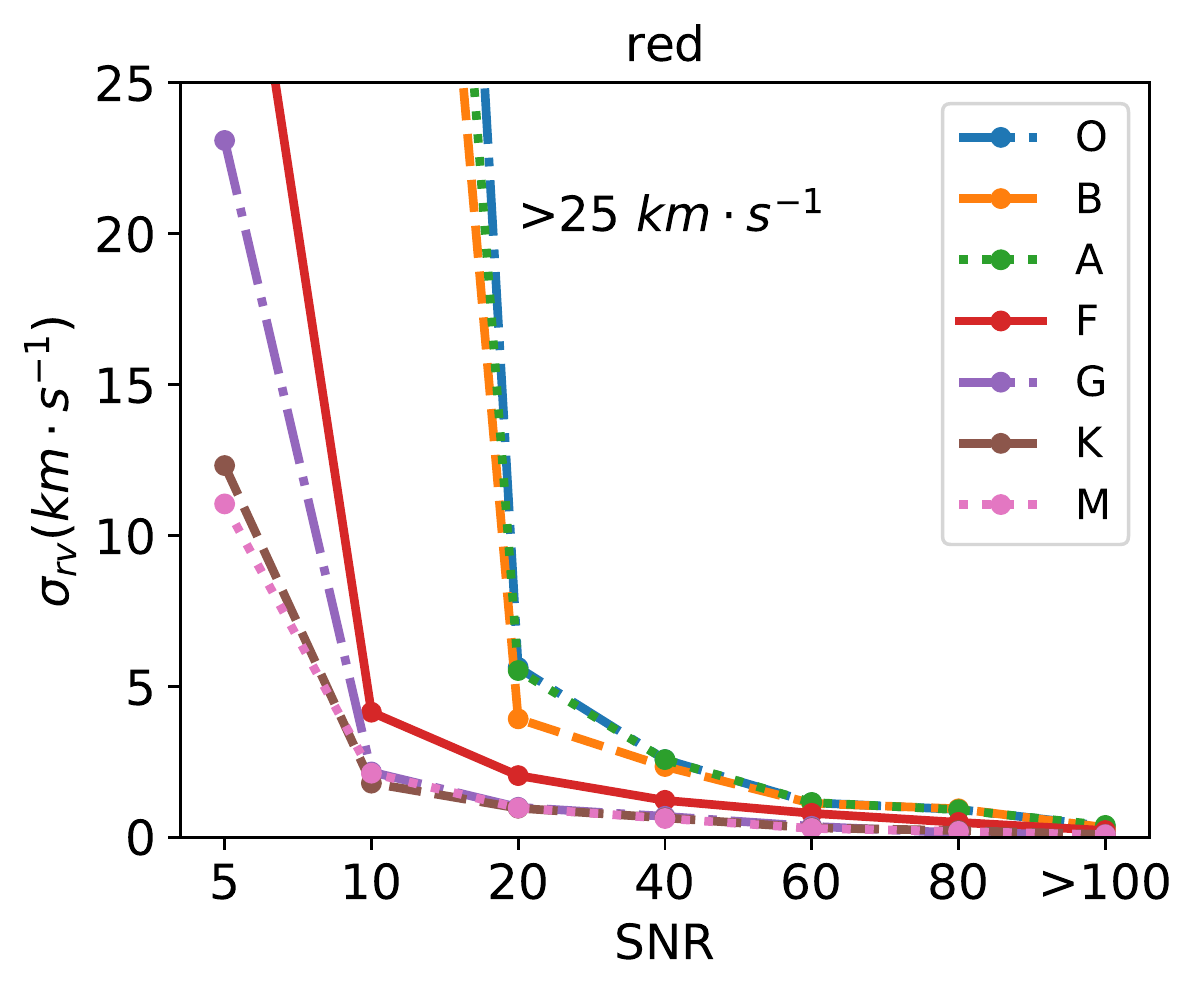}
    }
  \quad
  \subfigure[]{
    \includegraphics[height=6cm,width=7cm]{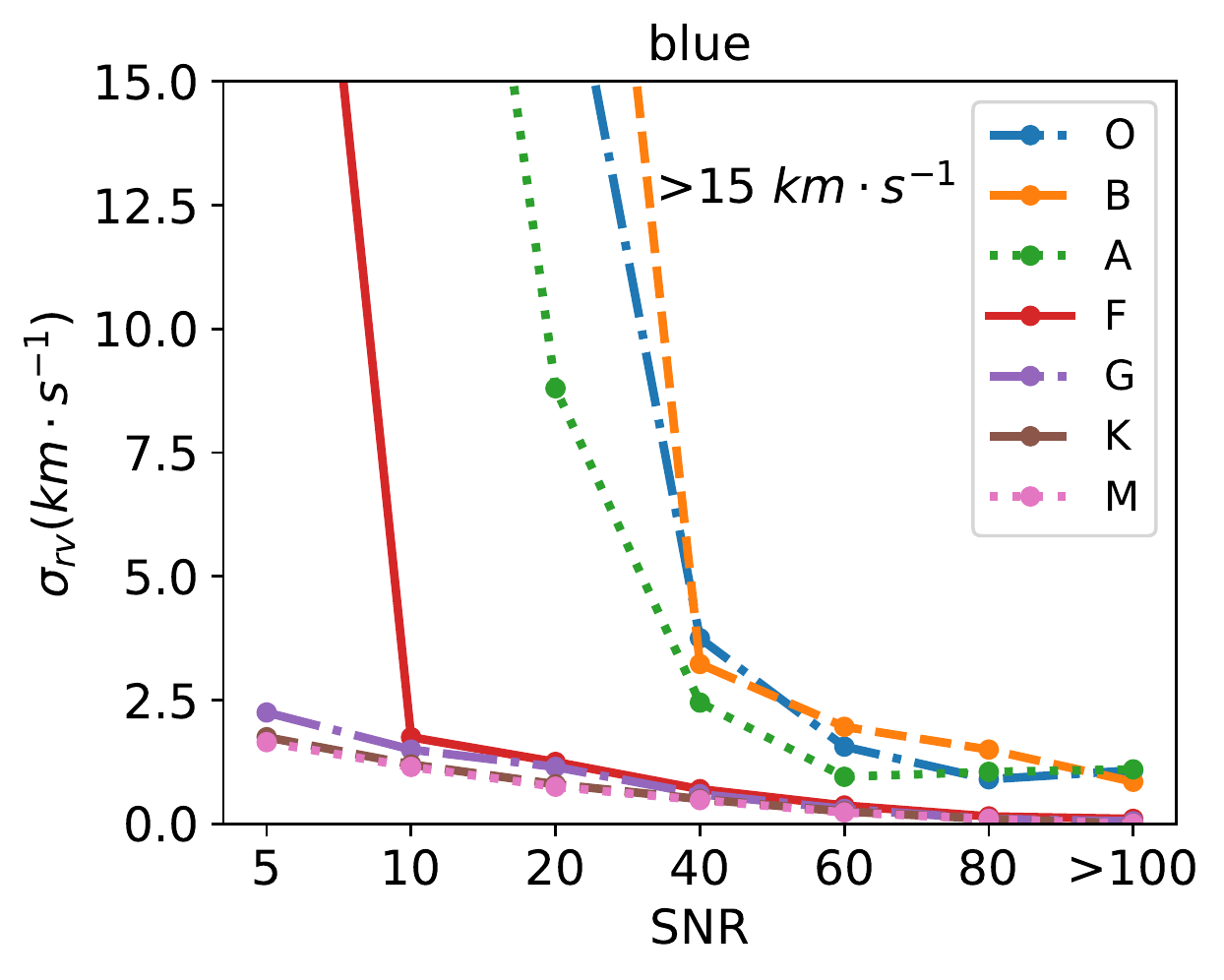}
    }
  \quad

  \caption{Panels (a) and (b) show the precision ($\sigma_{rV}$) of this relative radial velocity measurement method for different types spectra in different SNR for red and blue arms, respectively.}
  \label{fig:8}
  \end{figure*}

We conducted the test for single line spectra at first, we selected 7 spectra with different types with high SNR (SNR$>$100) in LAMOST MRS. For these 7 selected spectra, we simulated 1,000 spectra at different SNR for each type. For each SNR, we randomly selected a spectrum as the template to measure the relative radial velocity of them. And the ground true radial velocity of these spectra should be zero. For the different types of spectra in the varying SNR, we obtained the standard deviation ($\sigma_{RV}$) from Monte Carlo simulation results as the measurement precision. Figure~\ref{fig:8} (a) and (b) show the precision improves with the increasing SNR in the red and blue arm, respectively. We infer from figure~\ref{fig:8} that there is a relatively low precision of type O, B, and A when SNR less than 40, especially in the blue arm (see panel (b)). For the type O, B, and A, there are fewer lines to measure radial velocity in the blue arm and the signal is mixed up with noise, so there is a relatively low precision of type O, B, and A when SNR is less than 40 in the blue arm. This is improved a lot in the red arm, since $H_{\alpha}$ line is dominated in red arm. For type F, G, K, and M, we can derive the more accurate radial velocity in relatively high precision even though they are in a low SNR, because of their rich populations of spetral lines.

Then we test on pulsating variable stars. As a  pulsating variable star, its effective temperature, spectra feature, and radial velocity may change in a short period, such as RR-Lyrae. Because the periodic expansion and contraction of these stars cause a temperature change, and the luminosity also periodically increases and decreases, it appears that its brightness periodically brightened and darkened. It indicates that the spectra of the pulsating variable star may significantly change in a short period with its large changes in temperature.

\begin{figure*}[!t]
  \centering
  \subfigure[]{
    \includegraphics[height=6cm,width=7cm]{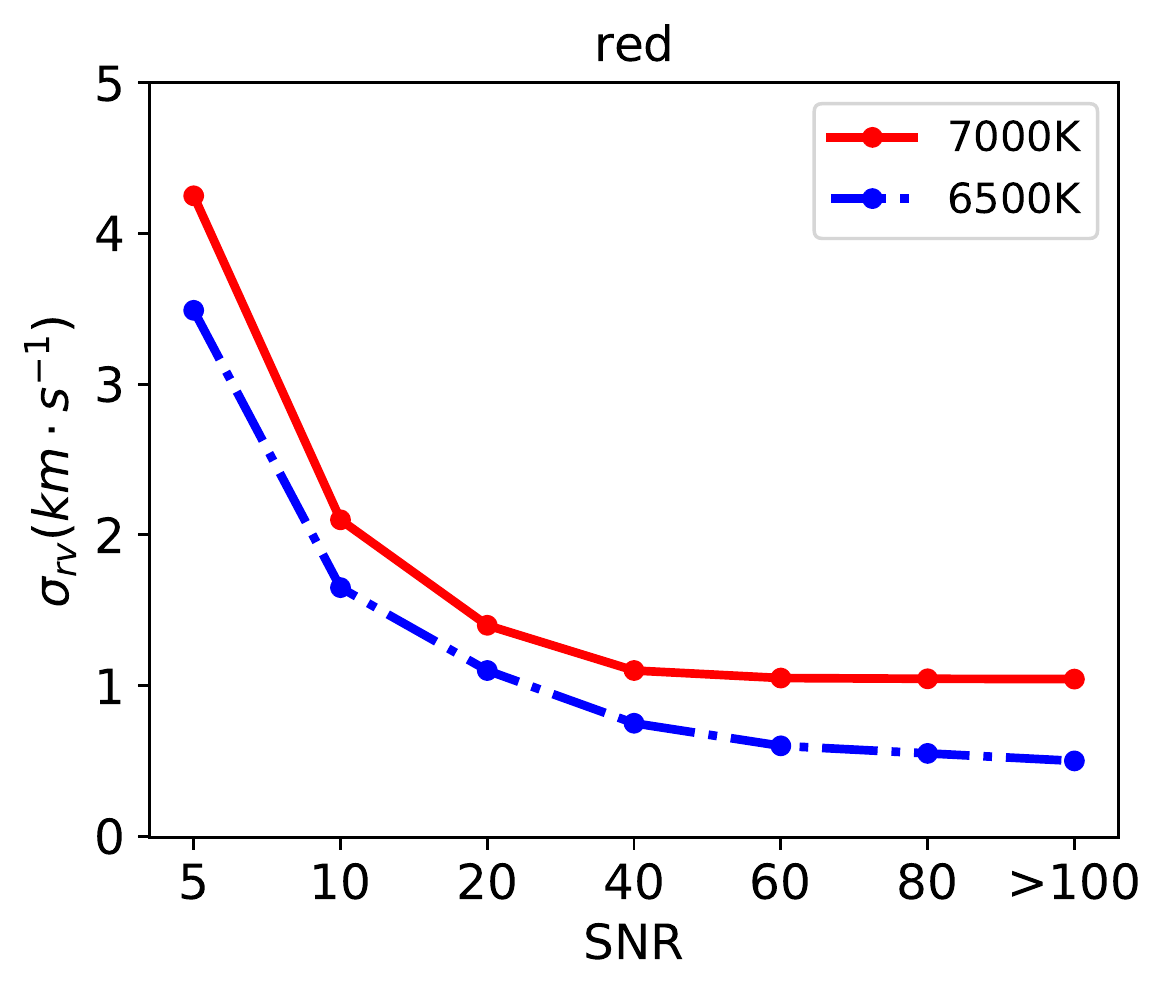}
    }
  \quad
  \subfigure[]{
    \includegraphics[height=6cm,width=7cm]{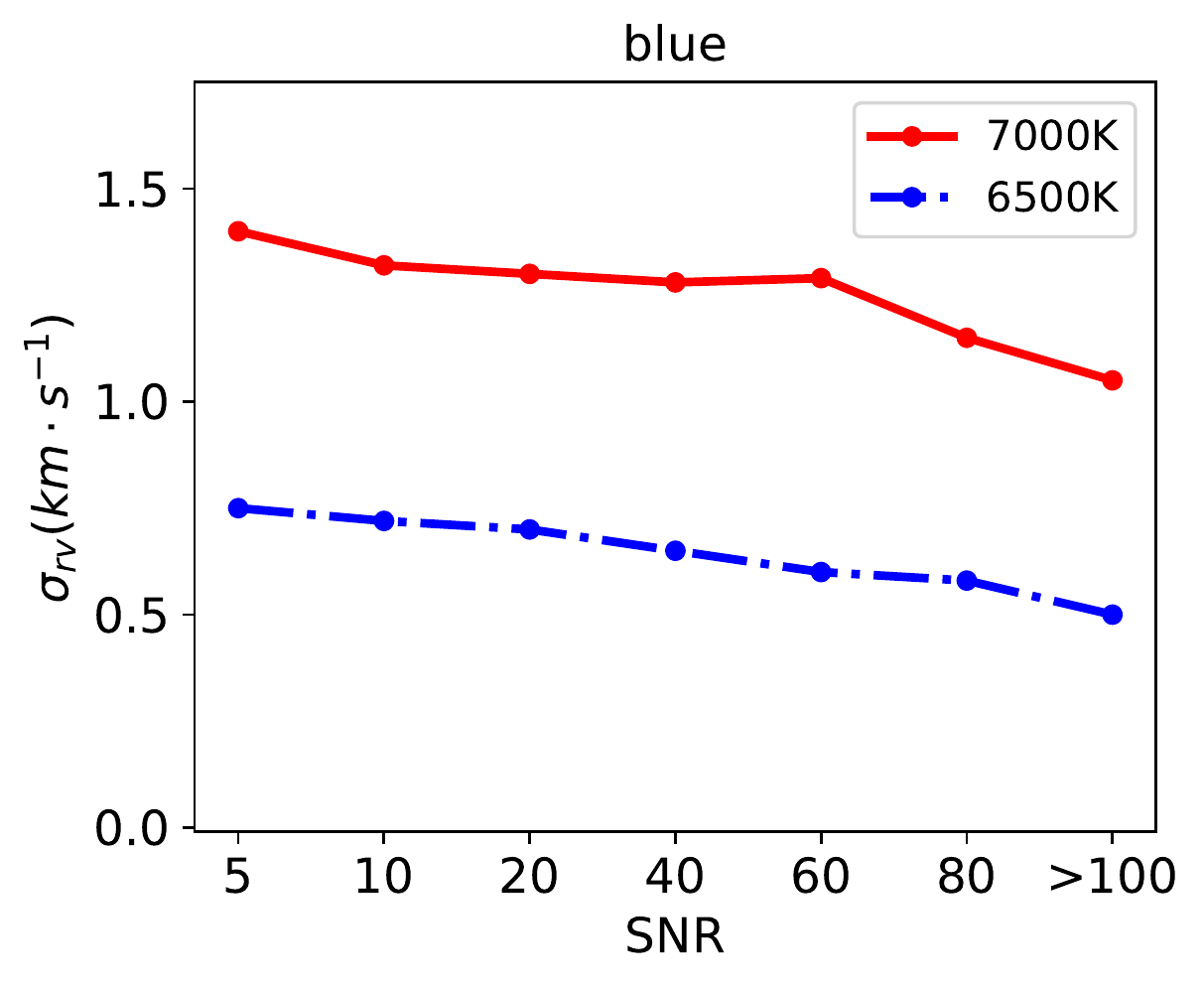}
    }
  
  \caption{Panels (a) and (b) show the relative radial velocity measurement precision($\sigma_{rv}$) for the stars with large temperature change in different SNR for red and blue arm, respectively.}
  \label{fig:9}
\end{figure*}

So, we mimic pulsating variable stars' spectra by using the standard template with different temperatures of 6000\,K, 6500\,K, 7000\,K, with fixed metallicity of -1 dex and logg of 2.5 dex. We assumed that the spectra in 3 temperatures are the observational spectra in the time-domain survey of an RR-Lyrae star. And we produced 1000 mock spectra by the different SNR for the temperatures in 6500\,K and 7000\,K, and measured their radial velocity by comparing with the spectrum with temperature in 6000\,K. Then we obtained the standard deviation from Monte Carlo simulation results as the measurement precision. Figure~\ref{fig:9} shows $\sigma_{RV}$ in different temperatures with changing SNR in the red and blue arms. In figure~\ref{fig:9}, the red solid line is the result of the temperature of 7000\,K and the blue dash-dotted line is for the spectrum with 6500\,K. It is seen that the precision is also increasing with the increasing SNR and is relatively higher for the smaller temperature change. Figure~\ref{fig:9} shows that even the pulsators change their temperature by 1000K, the relative radial velocities are still reliable.

For the binary systems, we can derive the differential radial velocity for single-line spectroscopic binary. But for double-line spectroscopic binary, we can derive the radial velocity with this method only if that we can obtain both of the spectra at the time of eclipse and non-eclipse. Because the basic idea for this radial velocity measurement method is to calculate the similarity for two spectra of the same target in different exposures. If the double-line spectroscopic binary are always observed at the time of non-eclipse, i.e. shows double-line in all exposures, the relative radial velocity with assumption of single star is valid anymore. So, for double-line spectroscopic binary, the differential radial velocity can not be measured accurately with this likelihood if the observational spectra are always in double-line. Their velocities should be dealt with in a different way but we can pick them out as the variable stars. Therefore, in this paper, the double-line binary stars are only marked and will be treated with other approach.

\section{Conclusion} \label{sec:Conclusion}
In this work, we mainly develop an approach in the manner of self-calibration by using differential radial velocity for LAMOST time-domain medium resolution search, and it is practicable to correct systematic shift in different exposures by using the differential radial velocities. The precision of radial velocity zero-point for different exposures after correcting systematic shift can reach below 0.5\,\kms. 

In total, 2,215,918 spectra from the 59 time-domain plates of LAMOST MRS of DR7 were calibrated in this work, including 1,170,445 of the red arm and 1,045,473 of the blue arm. And  the whole catalog of calibrated relative radial velocities of the spectra for the LAMOST time-domain medium resolution survey is provided in this paper. And after self-calibration in red and blue arms, we obtain the 65,123 constant stars candidates, whose radial velocity variation reach around zero after calibration. This self-calibration method can be used for the long-term observation of LAMOST time-domain medium resolution survey and can be used to select constant stars in a relatively high precision. These initial constant star candidates can provide a library of standard star candidates of radial velocity for the LAMOST time-domain medium resolution search. As the observation time increases, the library of the constant stars becomes more accurate. For the scientific significance, the variable stars with a relatively small amplitude can be detected and their radial velocities are calibrated effectively by this self-calibration method.

\begin{acknowledgements}
This work is supported by National Key R\&D Program of China No. 2019YFA0405500. C.L. Thanks the National Natural Science Foundation of China (NSFC) with grant No. 11835057, the State Natural Sciences Foundation
Monumental Projects with Nos. 12090040. Z.W.H, 12090041. H.W. Guoshoujing Telescope (the Large Sky Area Multi-Object Fiber Spectroscopic Telescope LAMOST) is a National Major Scientific Project built by the Chinese Academy of Sciences. Funding for the project has been provided by the National Development and Reform Commission. LAMOST is operated and managed by the National Astronomical Observatories, Chinese Academy of Sciences.

\end{acknowledgements}
  
\bibliographystyle{raa}
\bibliography{ms2021-0140}

\end{document}